\documentclass[aps,reprint,showpacs,floatfix,nofootinbib,superscriptaddress]{revtex4-1}
\usepackage[english]{babel}
\usepackage{graphicx}
\usepackage{dcolumn}
\usepackage{bm}
\usepackage{amssymb}
\usepackage{amsmath}
\usepackage{amsfonts}
\usepackage{subfig}
\begin{document}
\newcommand{\newc}{\newcommand}
\newc{\ra}{\rightarrow}
\newc{\lra}{\leftrightarrow}
\newc{\beq}{\begin{equation}}
\newc{\eeq}{\end{equation}}
\newc{\barr}{\begin{eqnarray}}
\newc{\earr}{\end{eqnarray}}

\title{Neutral Current Coherent Cross Sections- Implications on\\
Gaseous Spherical TPC's  for detecting SN and Earth neutrinos}

\author{J.~D.~Vergados}
\affiliation{Theoretical Physics Division, University of Ioannina, Ioannina, Gr 451 10, Greece}

\begin{abstract}
The detection of galactic supernova (SN) neutrinos represents one of the future frontiers of low-energy neutrino physics and astrophysics. The neutron coherence of neutral currents (NC) allows quite large cross sections in the case  of neutron rich targets, which can be exploited in detecting  earth and sky neutrinos by measuring nuclear recoils.  The
collapse of a neutron star liberates a gravitational binding energy of about $3\times10^{53}$ erg, 99$\%$ of
which is transferred to neutrinos and antineutrions  of all the flavors and only 1$\%$ to the kinetic energy of the explosion. In other words,
a core-collapse supernova represents one of the most powerful source of (anti)neutrinos in the Universe.  These (NC) cross sections are not dependent on flavor conversions and, thus, their measurement will provide useful information about the neutrino source. In particular  the case of SN they will yield information about the primary neutrino fluxes, i.e. before flavor conversions in neutrino sphere. The advantages of large gaseous low threshold and high resolution time projection counters (TPC) detectors  TPC detectors will be discussed. These are especially promising since they are expected to be relatively cheap and easy to maintain.  The information thus obtained can also be useful to  other flavor sensitive detectors, e.g. the large liquid scintillation detectors like LENA.   All together such detectors will provide invaluable information on the astrophysics of core-collapse explosion and on the neutrino mixing parameters. In particular, neutrino flavor transitions in SN envelope might be sensitive to the value of $\theta_{13}$ and to the unknown neutrino mass hierarchy. Till a real SN explosion is detected, one can use available earth neutrino sources with similar energy spectra to test the behavior of these detectors. Among them,  the ORNL Neutron Spallation source (SNS) and boosted radioactive neutrino beams are good candidates.

\end{abstract}



\maketitle
\section{Introduction}
The detection of galactic supernova (SN) neutrinos represents one of the future frontiers of low-energy neutrino physics and astrophysics.\footnote{see G. Raffelt, "Physics opportunities with supernova neutrinos"', Proceedings of the FIFTH SYMPOSIUM ON LARGE TPCs
FOR LOW ENERGY RARE EVENT DETECTION
and workshop on neutrinos from Supernovae.}
 In this paper we are going to discuss the relevant physics for the design and construction of a gaseous spherical TPC  for dedicated supernova detection, exploiting the coherent neutrino-nucleus elastic scattering due to the neutral current interaction. This detector can draw on the progress made in recent years in connection with measuring nuclear recoils in dark matter searches. It has low threshold and high resolution, it is relatively cheap and easy to maintain. Before doing this, however, we will briefly discuss the essential physics of neutrinos emitted in supernova explosions \cite{RAFFELT04} (for a review, see, e.g. the recent report \cite{JLMMM06})\footnote{ see also G. Fuller, "Neutrinos from Supernovae", Proceedings of the FIFTH SYMPOSIUM ON LARGE TPCs
FOR LOW ENERGY RARE EVENT DETECTION
and workshop on neutrinos from Supernovae.}.
\section{The primary supernova neutrino flux.}
We will assume that the neutrino spectrum can be described be a Fermi Dirac Distribution with a given temperature $T$ and a chemical potential $\mu=a T$. The constants $T$ and   $a$ will be treated as free parameters. Thus
\beq
f_{sp}(E_{\nu},T,a)={\cal N}\frac{1}{1+\exp{(E_{\nu}/T-a)}}
\eeq
where $\cal{N}$ is a normalization constant. The temperature $T$ is taken to be $3.5$, $5$ and $8$ MeV for electron neutrinos ($\nu_e$), electron antineutrinos ($\tilde{\nu}_e$) and all other flavors ($\nu_x$)  respectively. The parameter $a$ will be taken to be $0\le a \le 5$.
     \begin{figure}[!ht]
 \begin{center}
 \subfloat[]
{
\rotatebox{90}{\hspace{-0.0cm}{$f_{sp}\rightarrow$MeV$^{-1}$}}
\includegraphics[scale=0.7]{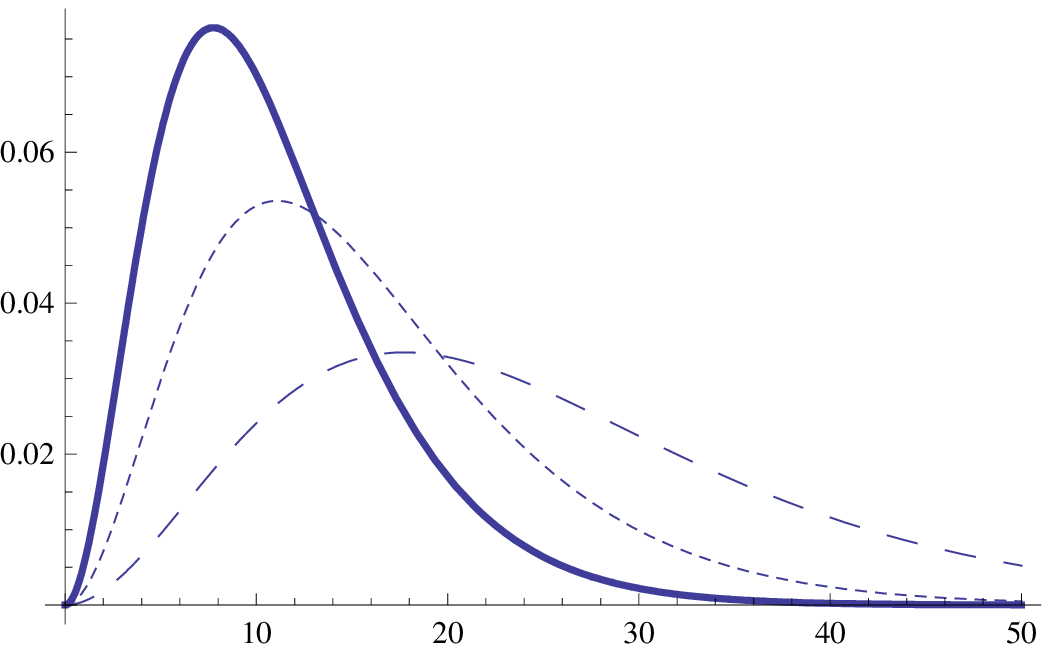}
}\\
 \subfloat[]
 {
\includegraphics[scale=0.7]{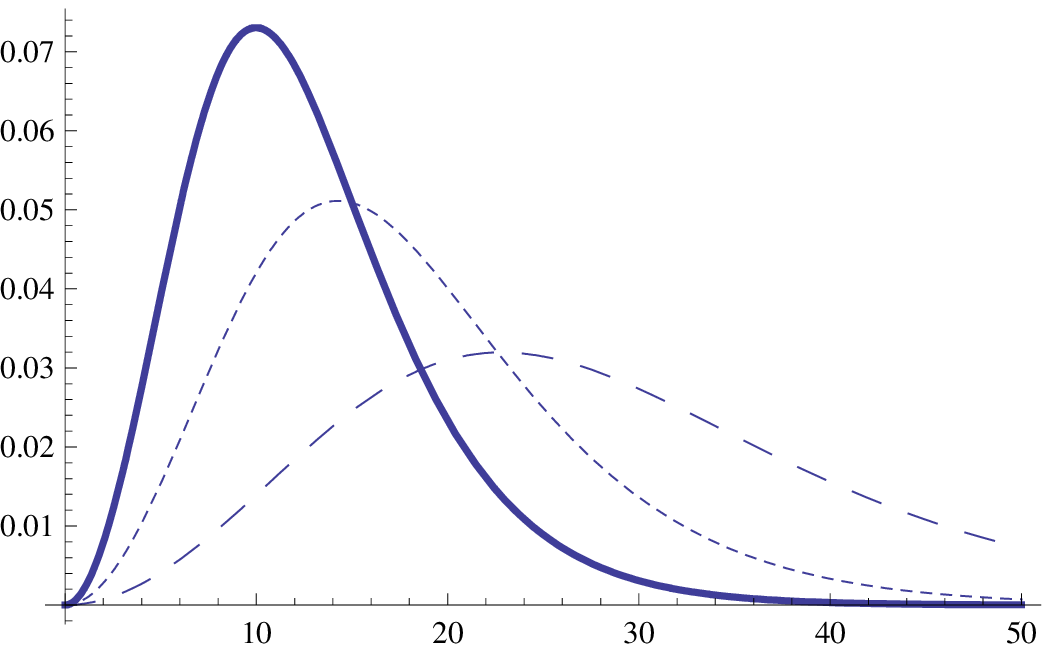}
}
\\
{\hspace{-0.0cm} {$E_{\nu}\rightarrow $MeV}}\\
 \caption{The normalized to unity SN spectrum for $a=0$ (a) and $a=3$  (b). The continuous, dotted and dashed curves correspond to $T$=3.5 ($\nu_e$), 5 ($\tilde{\nu}_e$) and 8 ($\nu_x$) respectively.}
 \label{fdis}
  \end{center}
  \end{figure}
  The average neutrino energies obtained from this distribution are shown in Table \ref{tab:avener}. 
  \begin{table}[t]
\caption{ The average supernova neutrino energies as a function of the parameters $a$ and $T$.
\label{tab:avener}
}
\begin{center}
\begin{tabular}{|c|c|c|c|}
\hline
$a$& \multicolumn{3}{c|}{$\prec E_{\nu}\succ$ (MeV)}\\
\hline
&$\nu_e$&$\tilde{\nu}_e$&$\sum_x\nu_x$\\
&T=3.5MeV&T=5MeV&T=8MeV\\
\hline
0& 11.0298 & 15.7569 & 25.211 \\
0.75& 11.4504 & 16.3578 & 26.1724 \\
 1.50&12.0787 & 17.2553 & 27.6085 \\
2.00& 12.6194 & 18.0277 & 28.8443 \\
 3.00&13.9733 & 19.9619 & 31.9391 \\
4.00& 15.6313 & 22.3305 & 35.7288 \\
 5.00&17.5179 & 25.0255 & 40.0408\\
 \hline
\end{tabular}
\end{center}
\end{table}
The number of emitted neutrinos \cite{RAFFELT04} can be obtained from the total emitted energy $U_{\nu}=3\times10^{53}$ erg
\beq
N_{\nu}=\frac{U_{\nu}}{\prec E_{\nu} \succ}.
\eeq
The obtained results are shown in Table \ref{tab:Nnu}
  \begin{table}[t]
\caption{ The number of primary neutrinos emitted in a typical supernova explosion  as a function of the parameters $a$ and $T$ in units of $10^{58}$.
\label{tab:Nnu}
}
\begin{center}
\begin{tabular}{|c|c|c|c|}
\hline
$a$& \multicolumn{3}{c|}{${N_{\nu}}/{10^{58}}$ }\\
\hline
&$\nu_e$&$\tilde{\nu}_e$&$\sum_x\nu_x$\\
&T=3.5 MeV&T=5 MeV&T=8 MeV\\
\hline
0& 0.282969 & 0.198079 & 0.495196 \\
0.75 &0.272575 & 0.190802 & 0.477006 \\
 1.50&0.258397 & 0.180878 & 0.452194 \\
 2.00&0.247326 & 0.173128 & 0.43282 \\
3.00& 0.223361 & 0.156353 & 0.390882 \\
 4.00&0.199669 & 0.139768 & 0.349421 \\
5.00& 0.178167 & 0.124717 & 0.311792\\
 \hline
\end{tabular}
\end{center}
\end{table}
The (time averaged) neutrino flux $\Phi_{\nu}=N_{\nu}/(4 \pi D^2) $ at a distance $D=10$ kpc=$3.1\times 10^{22}$cm is given in Table \ref{tab:Phinu}.
  \begin{table}[t]
\caption{ The (time integrated) neutrino flux, in units of $10^{12}$cm$^{-2}$, at a distance 10 kpc from the source.
\label{tab:Phinu}
}
\begin{center}
\begin{tabular}{|c|c|c|c|}
\hline
$a$& \multicolumn{3}{c|}{${\Phi_{\nu}}/{10^{12}\mbox{cm}^{-2}} $}\\
\hline
&$\nu_e$&$\tilde{\nu}_e$&$\sum_x\nu_x$\\
&T=3.5 MeV&T=5 MeV&T=8 Mev\\
\hline
0& 0.234318 & 0.164023 & 0.410057 \\
 0.75&0.225711 & 0.157997 & 0.394994 \\
1.50& 0.213971 & 0.149779 & 0.374448 \\
2.00& 0.204803 & 0.143362 & 0.358405 \\
 3.00&0.184958 & 0.129471 & 0.323677 \\
 4.00&0.16534 & 0.115738 & 0.289345 \\
 5.00&0.147534 & 0.103274 & 0.258185\\
 \hline
\end{tabular}
\end{center}
\end{table}
\section{The differential and total cross section}
\begin{figure}[!ht]
 \begin{center}
 \subfloat[]
 {
 \rotatebox{90}{\hspace{1.0cm} {$F^2(T_A) \rightarrow $}}
\includegraphics[scale=0.6]{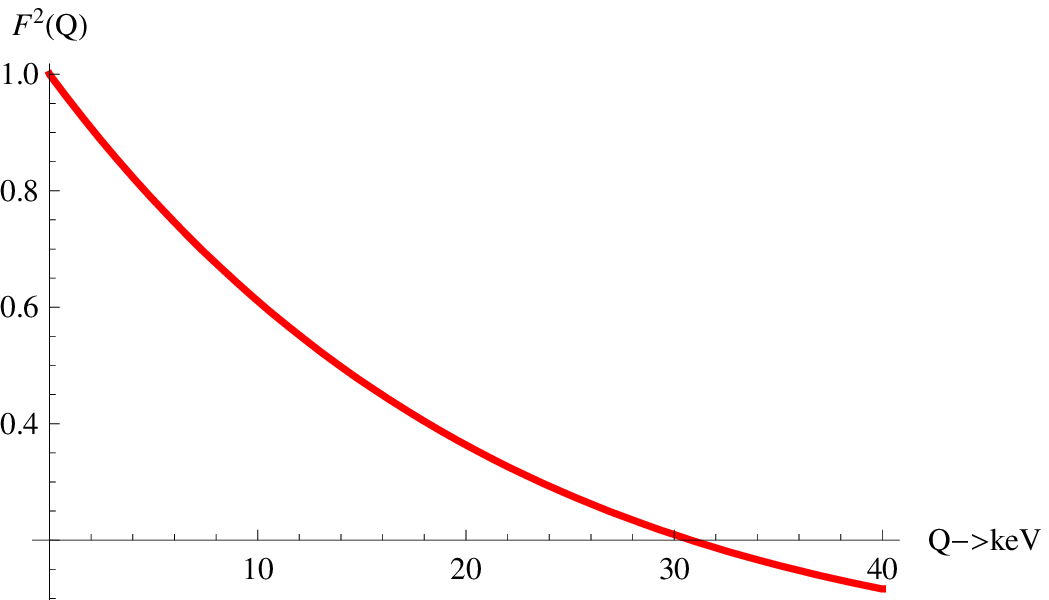}
}
\hspace{8.0cm}$T_A \rightarrow$ keV\\
\subfloat[]
{
\includegraphics[scale=0.6]{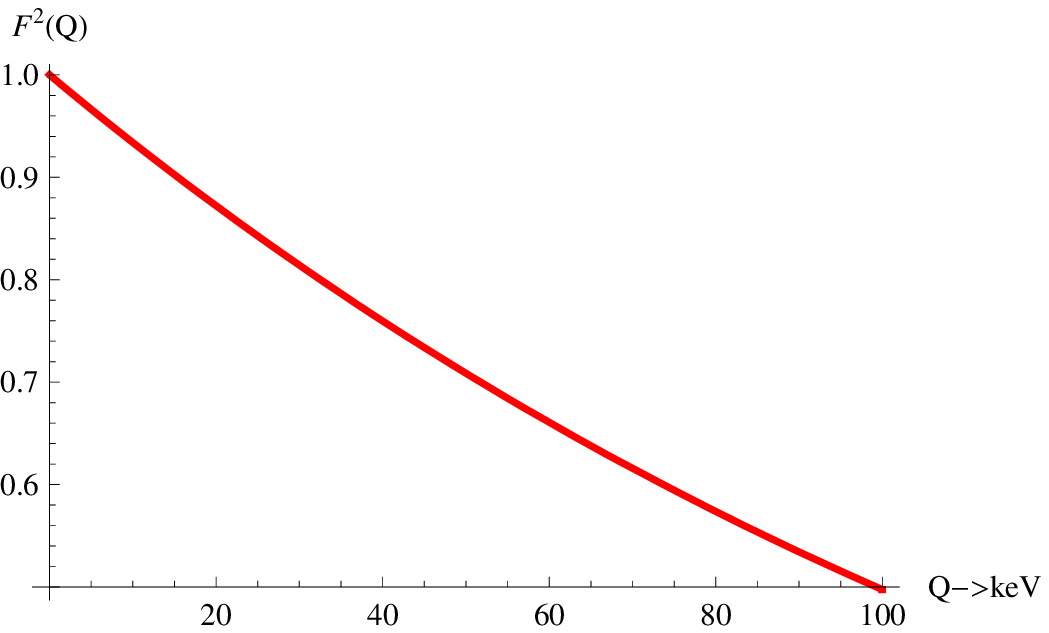}
}\\
\hspace{4.0cm}$T_A \rightarrow$ keV
 \caption{The square of the nuclear form factor, $F^2(T_A)$, as a function of the recoil energy for A=131 (a)
and A=40 (b). Note that the maximum recoil energy is different for each target.}
 \label{fig:ff}
 \end{center}
  \end{figure}

The differential cross  section for a given neutrino energy $E_{\nu}$ can be cast in the form\cite{VERGIOM06}:
\beq
 \left(\frac{d\sigma}{dT_A}\right)_{w}(T_A,E_{\nu})=\frac{G^2_F Am_N}{2 \pi}~(N^2/4) F_{coh}(T_A,E_{\nu}),
 \label{elaswAV1}
\eeq
with
\beq
F_{coh}(T_A,E_{\nu})= F^2(q^2)
  \left ( 1+(1-\frac{T_A}{E_{\nu}})^2
-\frac{Am_NT_A}{E^2_{\nu}} \right)
 \label{elaswAV2}
  \eeq
  where $N$ is the neutron number and $F(q^2)= F(T_A^2+2 A m_N T_A)$ is the nuclear form factor.
  The effect of the nuclear form factor depends on the target (see Fig. \ref{fig:ff}).
  
\begin{figure}[!ht]
 \begin{center}
 \rotatebox{90}{\hspace{0.0cm} $E^{\prime}_{th}\rightarrow $ keV}
\includegraphics[scale=0.7]{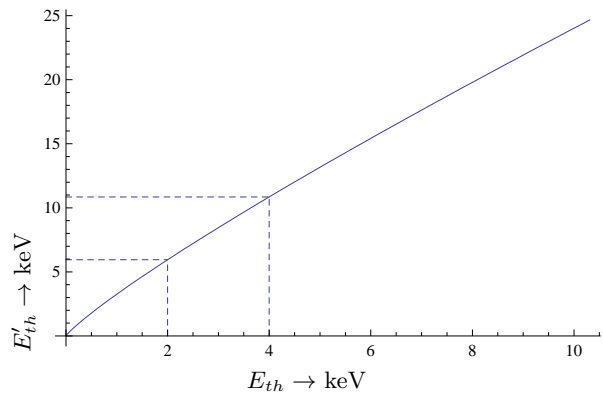}
\hspace{8.0cm}$E_{th}\rightarrow$ keV
 \caption{Due to quenching the threshold energy is shifted as shown in this figure, i.e. upwards from $E_{th}$ to $E^{\prime}_{th}$}
 \label{fig:qshift}
 \end{center}
  \end{figure}

   Since the SN   source is not "`monochromatic"' the above equation can be written as:
     \beq \frac{d\sigma}{dT_A}=\int_{E(T_A)}^{(E_{\nu})_{\mbox{\tiny{max}}}}\left(\frac{d\sigma}{dT_A}\right)_{w}(T_A,E_{\nu})f_{sp}(E_{\nu},T,a)d E_{\nu}
  \label{elaswAV3}
  \eeq
  Where $(E_{\nu})_{\mbox{\tiny{max}}}$ is the maximum neutrino energy and 
  $$ E(T_A)=\frac{T_A}{2}+\sqrt{\frac{T_A}{2}(M_A+\frac{T_A}{2})}$$
    Here $(E_{\nu})_{\mbox{\tiny{max}}}=\infty$.\\
    Integrating the total cross section of Fig. \ref{Fig:disigma.131} from  $T_A=E_{th}$ to infinity we obtain the total cross section.  The threshold energy $E_{th}$ depends on the detector.
    Furthermore for a real detector the nuclear recoil events are quenched, especially at low energies.
The quenching
factor for a given detector  is the ratio of the signal height for a recoil track divided by that of an electron signal with the same energy. We should not forget that the signal heights depend on the
velocity and how the signals are extracted experimentally. The actual quenching
  factors must be determined experimentally for each target. In the case of NaI the quenching
factor is 0.05, while for Ge and Si it is 0.2-0.3. For our purposes it is adequate, to multiply
the energy scale by an recoil energy dependent quenching factor, $ Q_{fac}(T_A)$
  adequately described by the Lidhard theory \cite{LIDHART}.  More specifically in our estimate of $Qu(T_A)$ we assumed a quenching factor of the following empirical form \cite{LIDHART}, \cite{SIMON03}:
\beq
Q_{fac}(T_A)=r_1\left[ \frac{T_A}{1keV}\right]^{r_2},~~r_1\simeq 0.256~~,~~r_2\simeq 0.153
\label{quench1}
\eeq
\begin{figure}[!ht]
 \begin{center}
 \subfloat[]
 {
 \rotatebox{90}{\hspace{0.0cm} $f_{q}(T_A)\rightarrow $}
\includegraphics[scale=0.6]{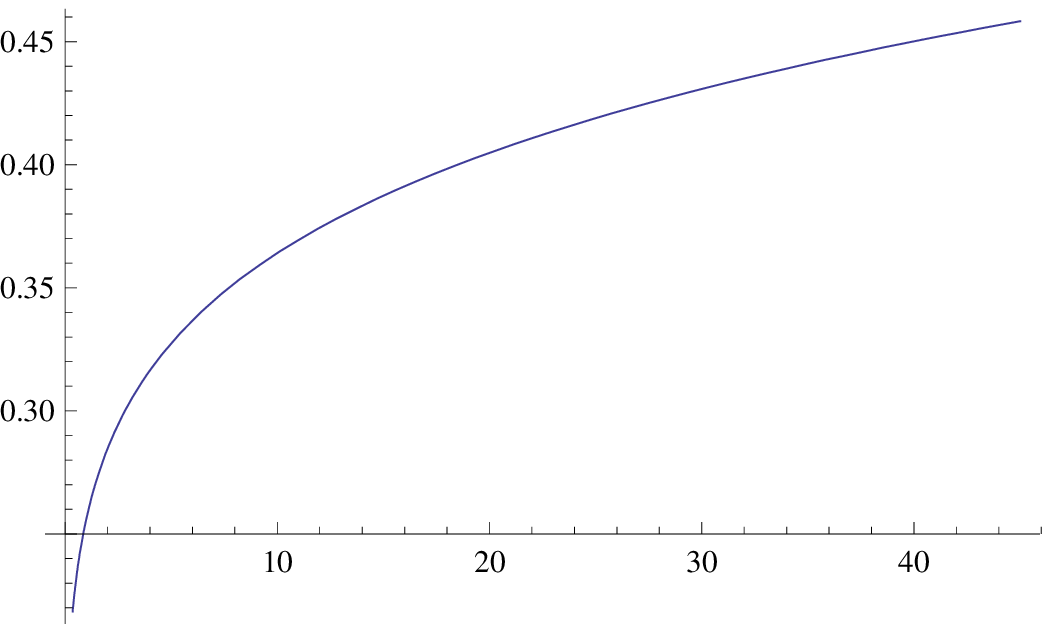}
}
\hspace{8.0cm}$T_A\rightarrow$ keV\\
\subfloat[]
{
\includegraphics[scale=0.6]{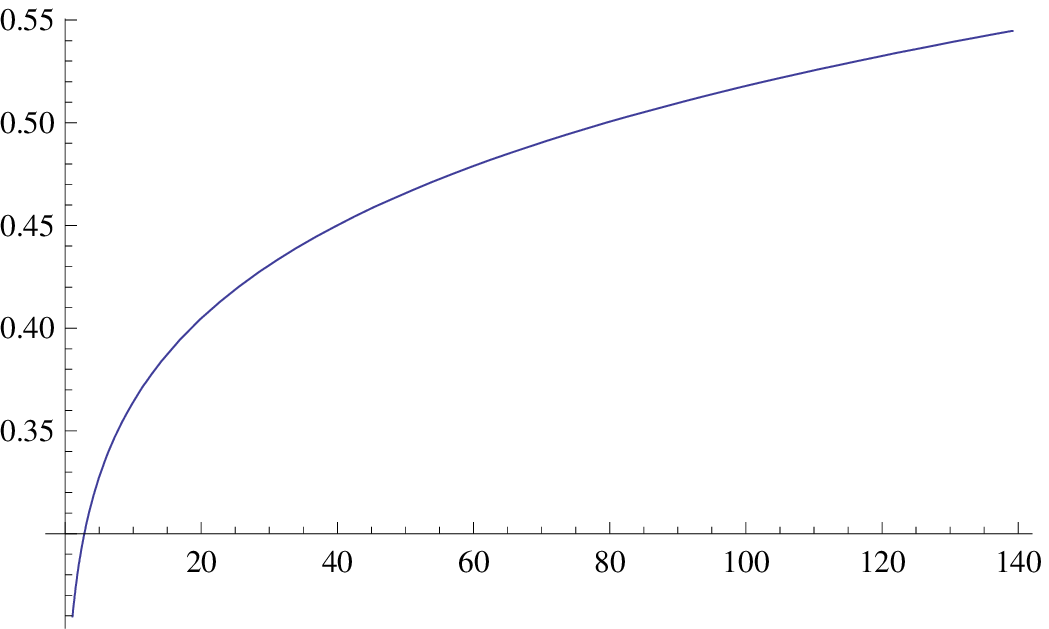}
}\\
\hspace{8.0cm}$T_A\rightarrow$ keV
 \caption{The quenching factor as a function of the recoil energy in the case of A=131 (a) and A=40 (b).}
 \label{fig:quench}
 \end{center}
  \end{figure}
The quenching factor, exhibited in Fig. \ref{fig:quench} for recoil energies appropriate for $^{131}$Xe
and $^{40}$Ar, were obtained assuming a quenching of the form of Eq. (\ref{quench1}).
In the presence of the quenching factor as given by Eq.( \ref{quench1})
the measured recoil energy is typically reduced by factors of about 3, when compared with the electron energy.  With the above quenching factor this relationship is shown in Fig. \ref{fig:qshift}. In other words
a threshold energy of electrons of $E_{th}=2$ keV becomes $E^{\prime}_{th}=6$  keV for nuclear recoils (see also Fig. \ref{Fig:disigma.131}).
 
  The number of the observed events for each neutrino species is found to be:
  \beq
  N_{ev}(a,T)=\Phi_{\nu} (a,T) \sigma({a,T,E_{th}}) N_{N}(P,T_0,R)
  \label{Tevent}
  \eeq
  \beq
   N_N(P,T_0,R)=\frac{P}{kT_0} \frac{4}{3} \pi R^3
  \eeq
  where $N_N$ is the number of nuclei in the target, which depends on the pressure, ($P$), the absolute temperature, ($T_0$) and the radius $R$ of the detector. We find:
  \beq
  N_N(P,T_0,R)=1.04\times10^{30} \frac{P}{10~\mbox{Atm}} \frac{300\,^0\mbox{K}}{T_0}\left (\frac{R}{10\mbox{m}} \right )^3
  \eeq
      \subsection{Results for the Xe target}
\begin{figure}[!ht]
 \begin{center}
 \subfloat[]
{
\rotatebox{90}{\hspace{-0.0cm}{$\frac{d \sigma}{d T_A}\rightarrow 10^{-39}$}cm$^2/$keV}
\includegraphics[scale=0.7]{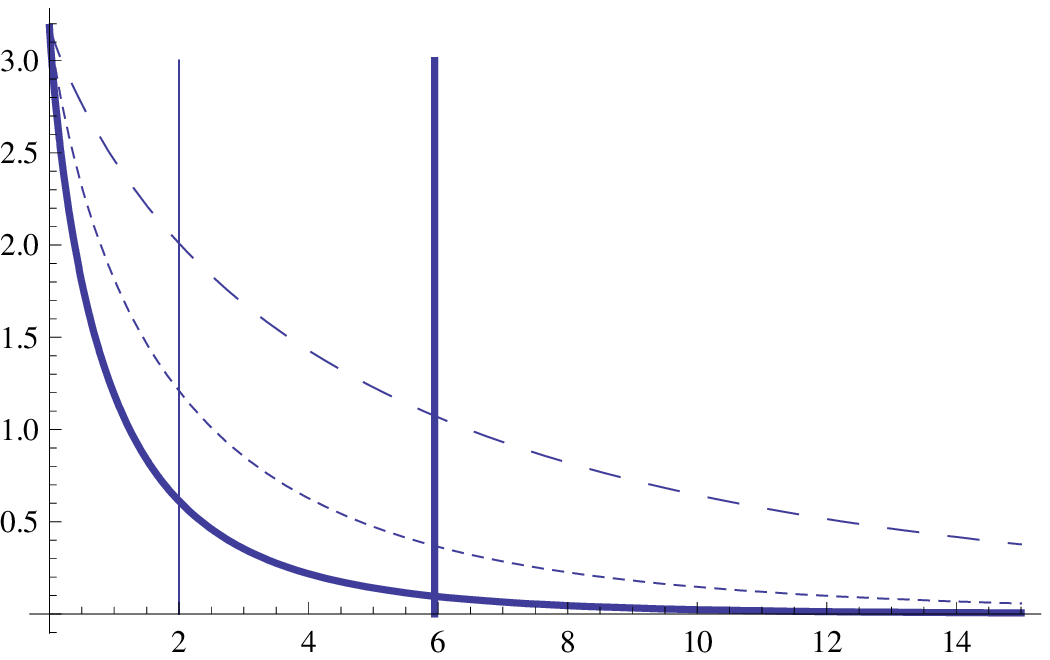}
}\\
 \subfloat[]
 {
\includegraphics[scale=0.7]{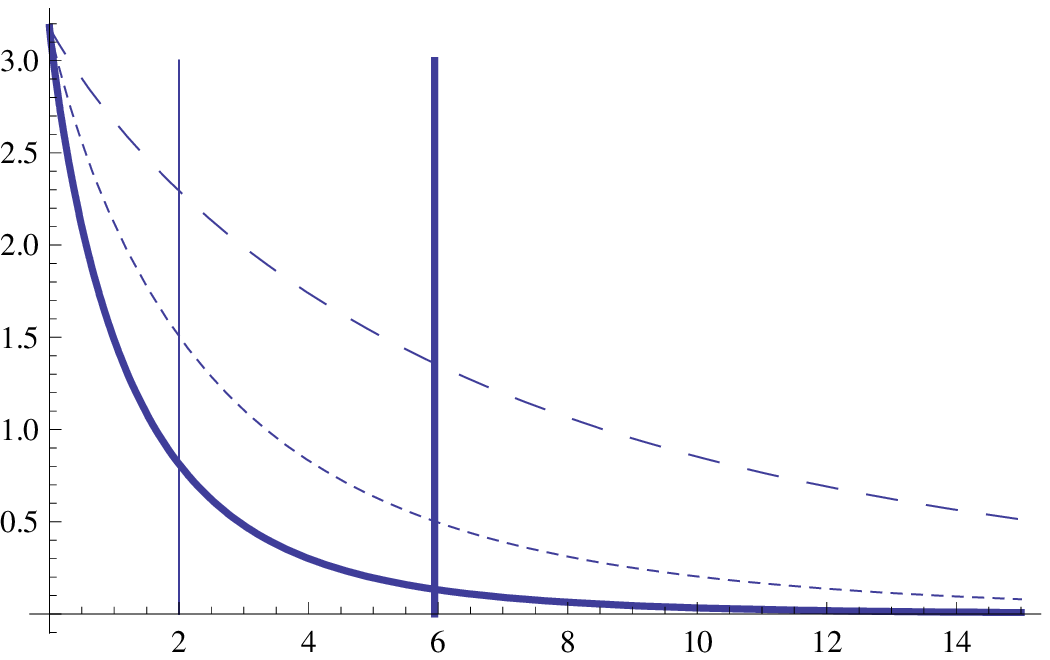}
}
\\
{\hspace{-0.0cm} {$T_{A}\rightarrow $keV}}\\
 \caption{The differential cross section for elastic neutrino nucleus scattering in the case of the target $^{131}_{54}$Xe as a function of the recoil energy $T_A$ in keV. In the case of a threshold energy of 2 keV, only the space on the right of the vertical bar is available. In the presence of quenching (see below) only the space on the right of the thick vertical bar is available. Otherwise the notation is the same as in Fig. \ref{fdis}}
 \label{Fig:disigma.131}
  \end{center}
  \end{figure}

\begin{figure}[!ht]
 \begin{center}
 \subfloat[]
{
\rotatebox{90}{\hspace{-0.0cm}{$\frac{\sigma(E_{th})}{\sigma(0)}\rightarrow$}}
\includegraphics[scale=0.7]{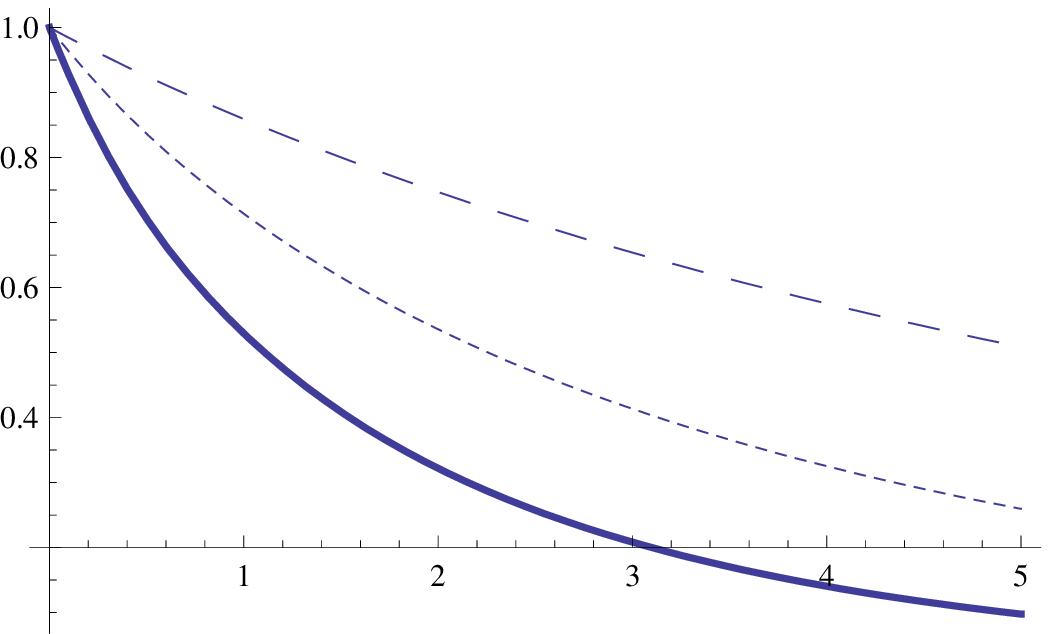}
}\\
 \subfloat[]
 {
\includegraphics[scale=0.7]{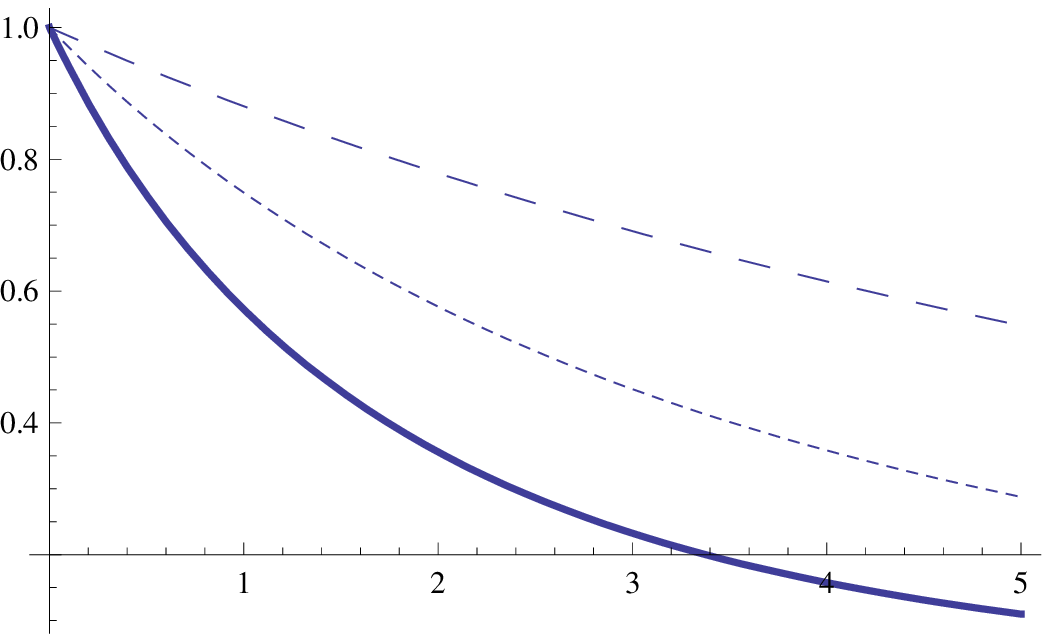}
}
\\
{\hspace{-0.0cm} {$E_{th}\rightarrow $keV}}\\
 \caption{The ratio of the cross section at threshold $E_{th}$ divided by that at zero threshold as a function of the threshold energy in keV in the case of a Xe target.
  Otherwise the notation is the same as in Fig. \ref{fdis}}
 \label{Fig:tsigma.131}
  \end{center}
  \end{figure}

\begin{figure}[!ht]
 \begin{center}
 \subfloat[]
{
\rotatebox{90}{\hspace{-0.0cm}{$\frac{\sigma(E_{th})}{\sigma(0)}\rightarrow$}}
\includegraphics[scale=0.7]{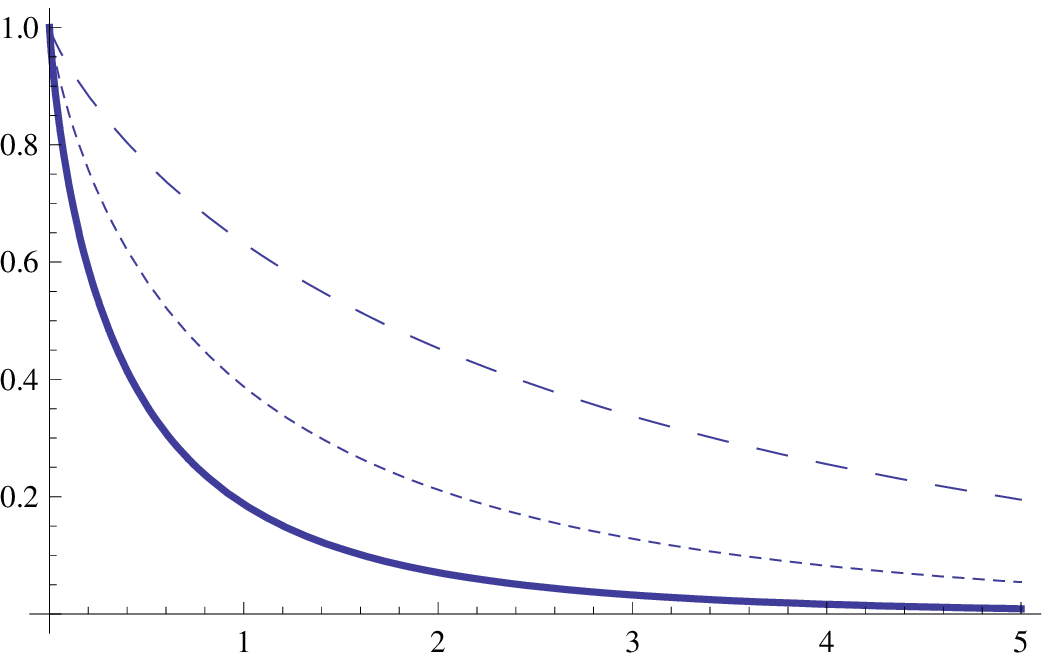}
}\\
 \subfloat[]
 {
\includegraphics[scale=0.7]{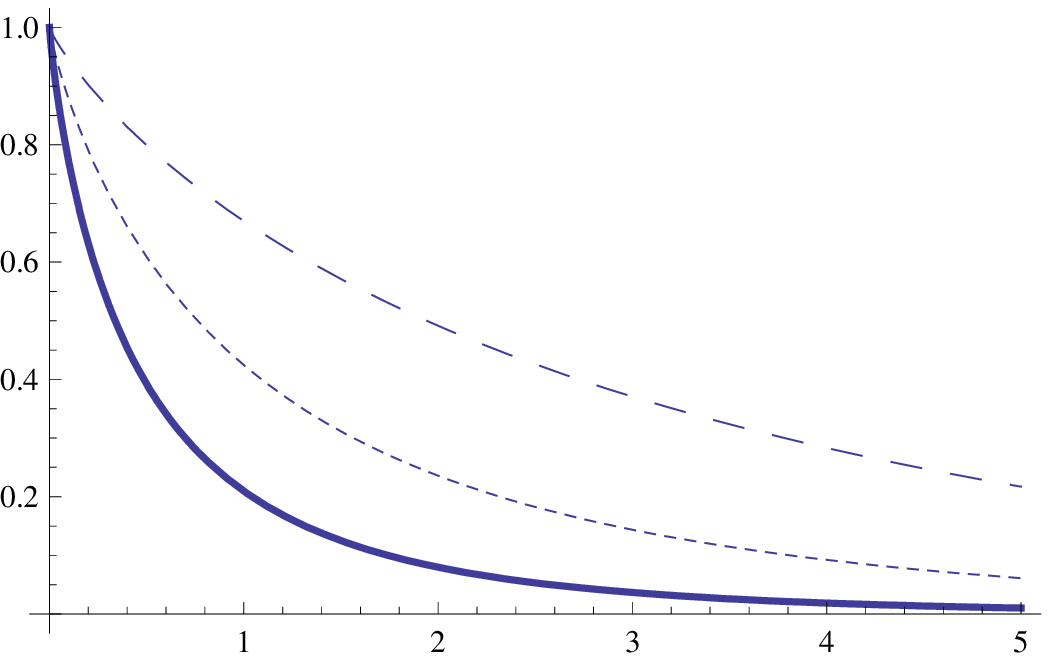}
}
\\
{\hspace{-0.0cm} {$E_{th}\rightarrow $keV}}\\
 \caption{
  The same as in Fig. \ref{Fig:tsigma.131}} taking into account the effect of quenching.
 \label{Fig:qtsigma.131}
  \end{center}
  \end{figure}

The differential cross section for neutrino elastic scattering, obtained with the above neutrino spectrum, on the target $^{131}_{54}$Xe is shown in Fig. \ref{Fig:disigma.131}.
  Integrating the total cross section of Fig. \ref{Fig:disigma.131} from  $T_A=0$ to infinity we obtain the total cross section given in table \ref{tab:sigma.131}.
   \begin{table}[t]
\caption{ The total neutrino nucleus cross section in the case of Xe target in units of $10^{-39}$cm$^{2}$ assuming zero detector threshold. 
\label{tab:sigma.131}
}
\begin{center}
\begin{tabular}{|c|c|c|c|c|}
\hline
$a$& \multicolumn{4}{c|}{${\sigma}/{10^{-39}\mbox{cm}^{2}} $}\\
\hline
&$\nu_e$&$\tilde{\nu}_e$&$\sum_x\nu_x$&Total\\
&T=3.5 MeV&T=5 MeV&T=8 MeV&\\
\hline
 0 & 4.117 & 8.312 & 19.764&32.194 \\
 0.75 & 4.361 & 8.815 & 20.921&34.097 \\
 1.50 & 4.749 & 9.608 & 22.727& 37.083\\
 2.00 & 5.104 & 10.330 & 24.346&39.780 \\
 3.00 & 6.074 & 12.288 & 28.621&46.083 \\
 4.00 & 7.408 & 14.966 & 34.147&56.521 \\
 5.00 & 9.118 & 18.364& 40.546&68.028\\
 \hline
\end{tabular}
\end{center}
\end{table}
The above results refer to an ideal detector operating down to zero energy threshold. In the case of non zero threshold the event rate is suppressed as shown in Fig.  \ref{Fig:tsigma.131}. 
 The total cross section is, of course, also affected by the quenching factor. This is exhibited in Fig. \ref{Fig:qtsigma.131}. The effect of quenching is quite important.
  
       Using Eq. \ref{Tevent} and the above total cross sections, after summing over all neutrino species (i.e. over all T),  we obtain the number of events shown in Table \ref{tab:rate.131}.
     \begin{table}[t]
\caption{ The total event rate as a function of $a$ in the case of a gaseous Xe target, under a pressure of 10 Atm and temperature 300 $^0$K, summed over all neutrino species assuming zero detector threshold. 
\label{tab:rate.131}
}
\begin{center}
\begin{tabular}{|r|r|r|}
\hline
a&R=10m&R=4m\\
\hline
 0 & 10872 & 696 \\
 0.75 & 11089 & 710 \\
 1.50 & 11427 & 731 \\
 2.00 & 11726 & 750 \\
 3.00 & 12483 & 799 \\
 4.00 & 13378 & 856 \\
 5.00 & 14288 & 914\\
 \hline
\end{tabular}
\end{center}
\end{table}
In the presence of a detector threshold of even 1 keV the above rates are reduced by about 20$\%$ (50$\%$) in the absence (presence) of quenching.
\subsection{The Ar target}
\begin{figure}[!ht]
 \begin{center}
 \subfloat[]
{
\rotatebox{90}{\hspace{-0.0cm}{$\frac{d \sigma}{d T_A}\rightarrow 10^{-40}$}cm$^2/keV$}
\includegraphics[scale=0.6]{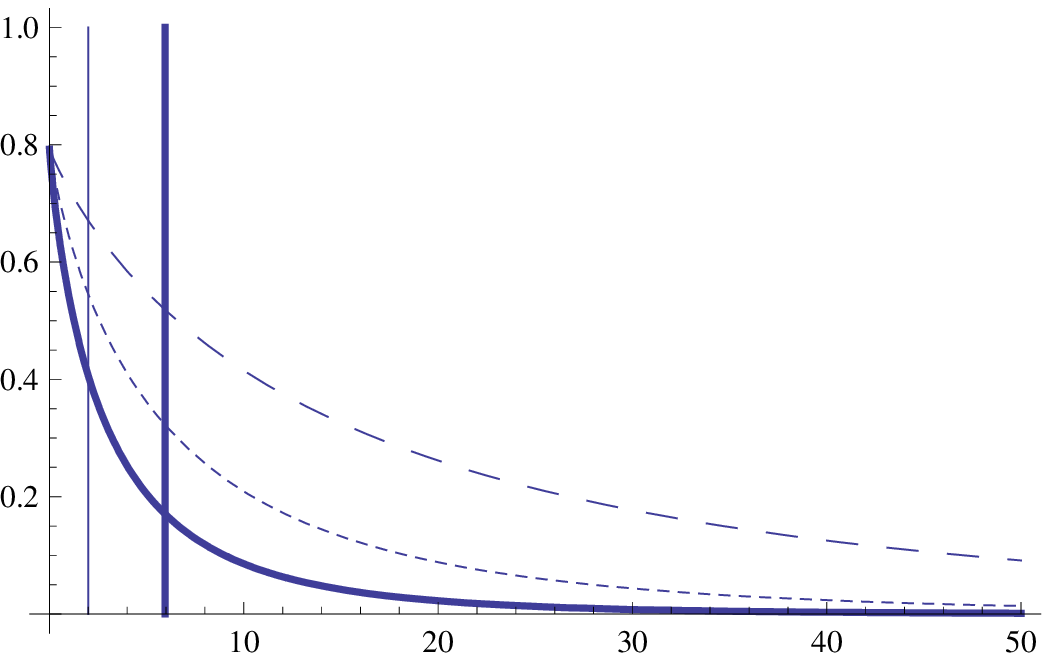}
}\\
 \subfloat[]
 {
\includegraphics[scale=0.6]{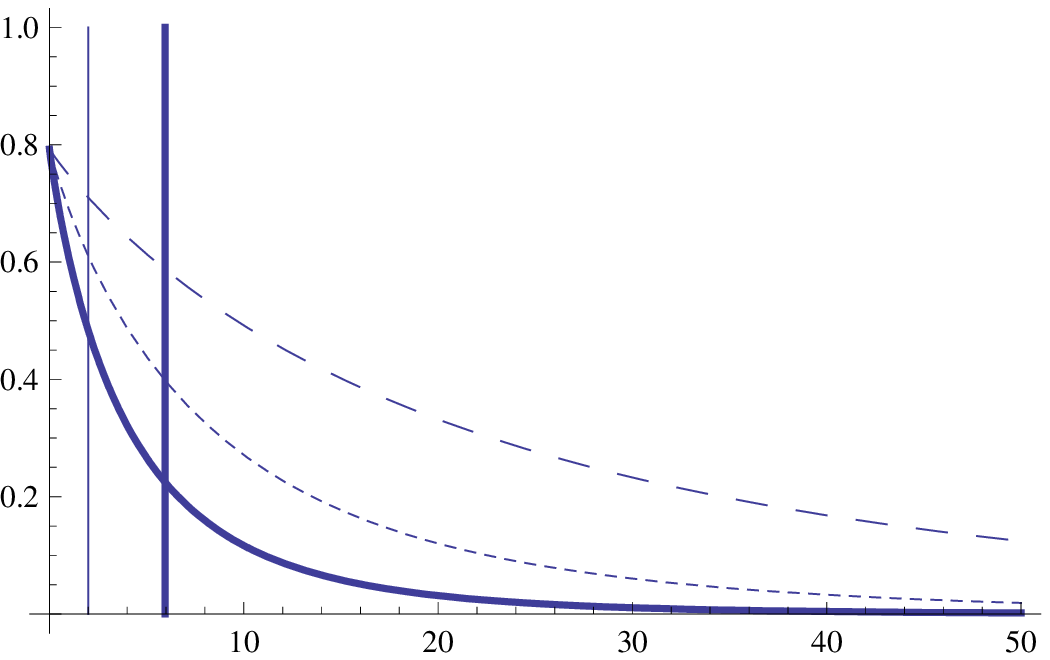}
}\\
{\hspace{-0.0cm} {$T_{A}\rightarrow $keV}}
 \caption{The same as in \ref{Fig:disigma.131} in the case of the Ar target.
 }
 \label{Fig:disigma.40}
  \end{center}
  \end{figure}

The differential cross for neutrino elastic scattering on the target $^{40}_{18}$Ar is shown in Fig. \ref{Fig:disigma.40}. For comparison we are currently calculating the differential cross sections to the excited states of $^{40}$Ar due to the neutral current. We also are going to calculate the charged current cross sections $(\nu_e,e^{-})$ and ($\tilde{\nu}_e,e^{+})$ on $^{40}$Ar, which are of interest in the proposal GLACIER, one of the large detectors\footnote{V. Tsakstara, private communication }.
  
   \begin{figure}[!ht]
 \begin{center}
 \subfloat[]
{
\rotatebox{90}{\hspace{-0.0cm}{$\frac{\sigma(E_{th})}{\sigma(0)}\rightarrow$}}
\includegraphics[scale=0.6]{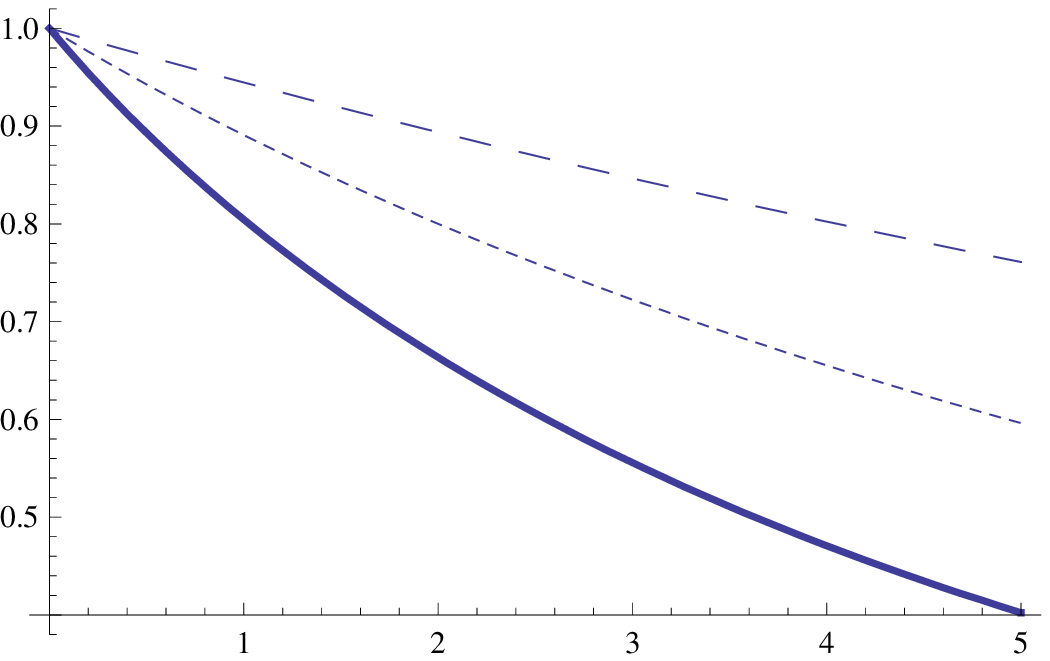}
}\\
 \subfloat[]
 {
\includegraphics[scale=0.6]{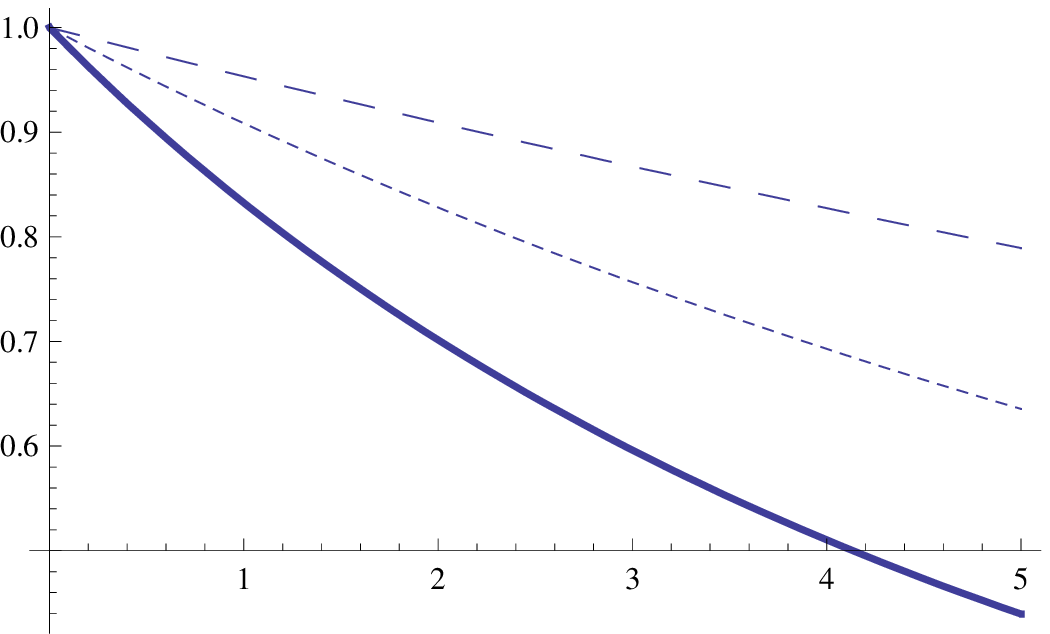}
}\\
{\hspace{-0.0cm} {$E_{th}\rightarrow $keV}}
 \caption{The same as in Fig. \ref{Fig:tsigma.131} in the case of the Ar target.}
 \label{Fig:tsigma.40}
  \end{center}
  \end{figure}
\begin{table}[t]
\caption{ The total neutrino nucleus cross section in the case of Ar target in units of $10^{-40}$cm$^{2}$ assuming zero detector threshold. 
\label{tab:sigma.40}
}
\begin{center}
\begin{tabular}{|c|c|c|c|c|}
\hline
$a$& \multicolumn{4}{c|}{${\sigma}/{10^{-40}\mbox{cm}^{2}} $}\\
\hline
&$\nu_e$&$\tilde{\nu}_e$&$\sum_x\nu_x$&Total\\
&T=3.5 MeV&T=5 MeV&T=8 MeV&\\
\hline
0 & 3.324 & 6.520 & 13.678 &23.521\\
 0.75 & 3.525 & 6.908 & 14.412&24.845 \\
 1.50& 3.843 & 7.518 & 15.528 &26.888\\
 2.00 & 4.133 & 8.067 & 16.497 &28.693\\
 3.00 & 4.917 & 9.537 & 18.905&33.359 \\
 4.00 & 5.990 & 11.488 & 21.690&39.168 \\
 5.00 & 7.353 & 13.843 & 24.480&45.676\\
 \hline
\end{tabular}
\end{center}
\end{table}
   \begin{figure}[!ht]
 \begin{center}
 \subfloat[]
{
\rotatebox{90}{\hspace{-0.0cm}{$\frac{\sigma(E_{th})}{\sigma(0)}\rightarrow$}}
\includegraphics[scale=0.6]{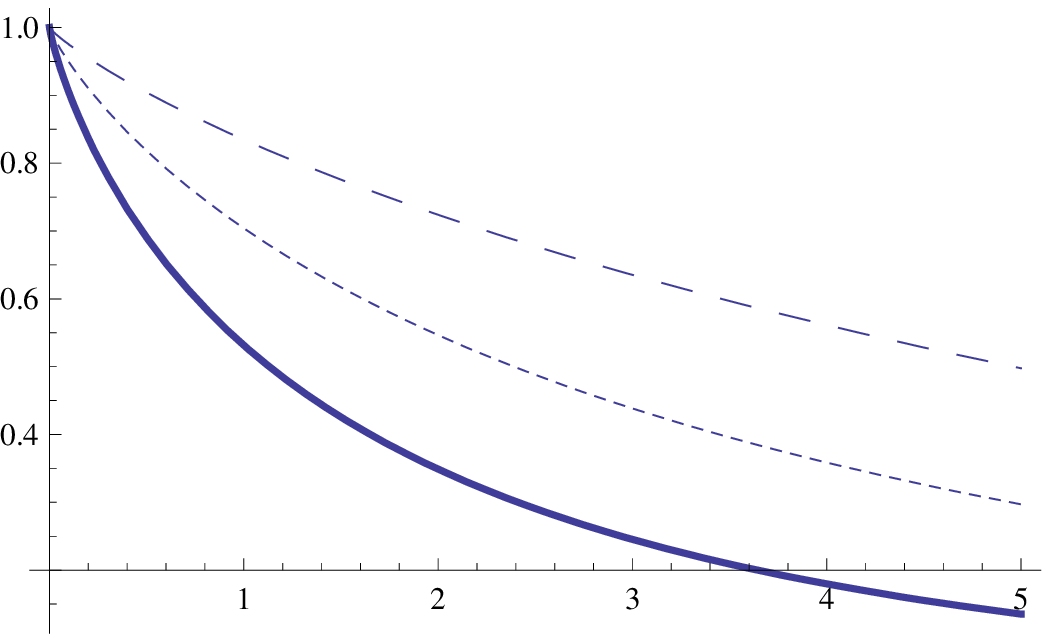}
}\\
 \subfloat[]
 {
\includegraphics[scale=0.6]{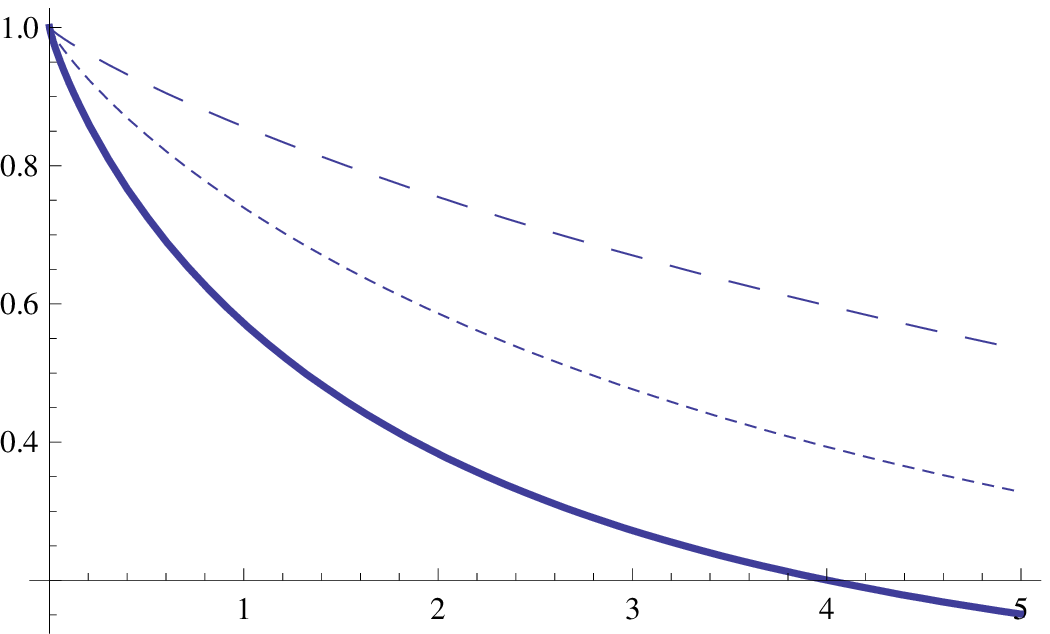}
}
\\
{\hspace{-0.0cm} {$E_{th}\rightarrow $keV}}
 \caption{The same as in Fig. \ref{Fig:qtsigma.131} in the case of the Ar target.}
 \label{Fig:qtsigma.40}
  \end{center}
  \end{figure}
   \begin{table}[t]
\caption{ The same as in Table \ref{tab:rate.131} in the case of the target Ar. We also show the corresponding events for 50 kton detector.}
\label{tab:rate.40}
\begin{center}
\begin{tabular}{|r|r|r|r|r|}
\hline
a&R=10m&R=4m&50 kton\\
&10 Atm &10Atm & (coh)\\
\hline
 0 & 193 & 12&2.8 $\times 10^{4}$ \\
 0.75 & 197 & 13&2.8 $\times 10^{4}$\\
 1.50 & 203 & 13&2.9$\times 10^{4}$ \\
 2.00 & 209 & 13&3.0 $\times 10^{4}$\\
 3.00 & 223 & 14 &3.2$\times 10^{4}$\\
 4.00 & 242 & 15 &3.5$\times 10^{4}$\\
 5.00 & 262 & 17&3.8$\times 10^{4}$\\
  \hline
\end{tabular}
\end{center}
\end{table}
In the presence of a detector threshold of even 1 keV the above rates are reduced by about 10$\%$ (30$\%$) in the absence (presence) of quenching.
\subsection{The Ne target}
\begin{figure}[!ht]
 \begin{center}
 \subfloat[]
{
\rotatebox{90}{\hspace{-0.0cm}{$\frac{d \sigma}{d T_A}\rightarrow 10^{-41}$}cm$^2/keV$}
\includegraphics[scale=0.7]{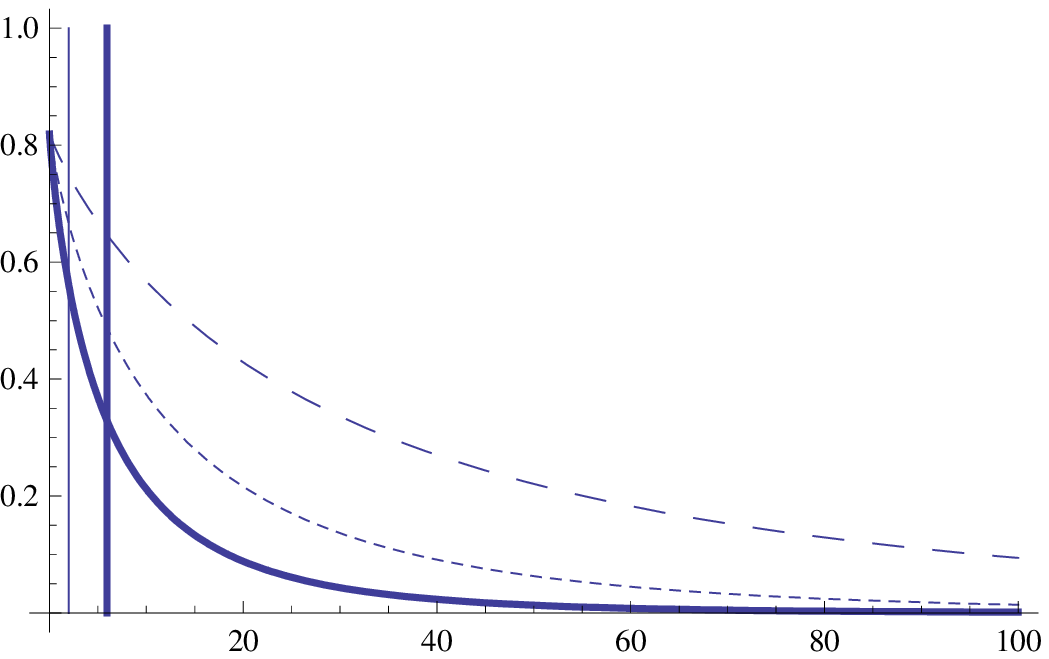}
}\\
 \subfloat[]
 {
\includegraphics[scale=0.7]{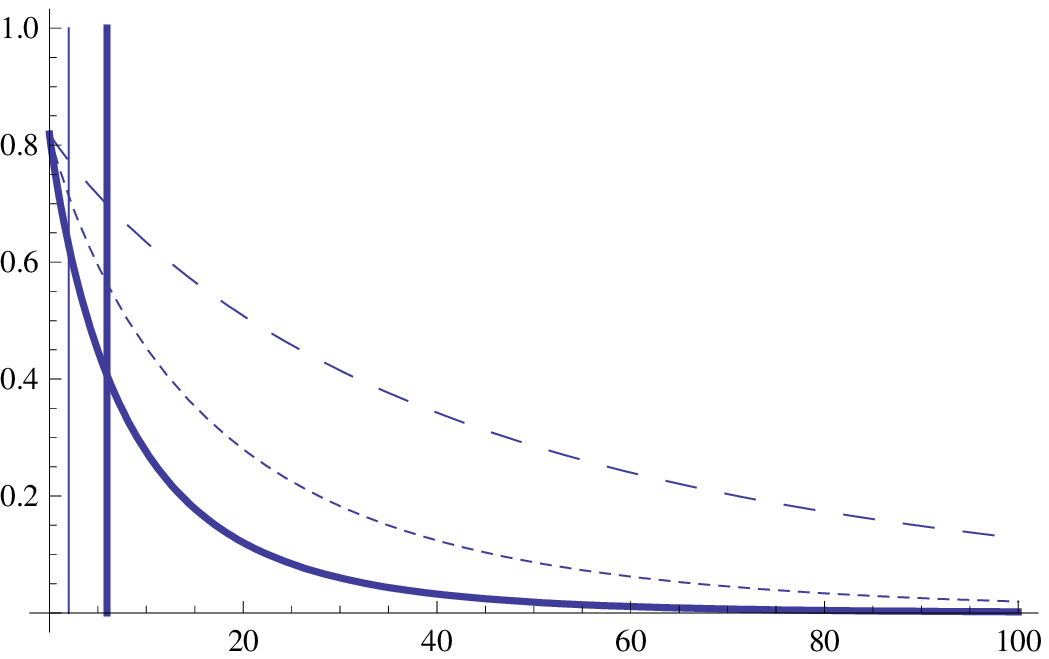}
}\\
{\hspace{-0.0cm} {$T_{A}\rightarrow $keV}}
 \caption{The same as in \ref{Fig:disigma.131} in the case of the Ne target.
 }
 \label{Fig:disigma.20}
  \end{center}
  \end{figure}

The differential cross for neutrino elastic scattering on the target $^{20}_{10}$Ne is shown in Fig. \ref{Fig:disigma.20}.
   \begin{table}[t]
\caption{ The total neutrino nucleus cross section in the case of Ne target in units of $10^{-41}$cm$^{2}$ assuming zero detector threshold. 
\label{tab:sigma.20}
}
\begin{center}
\begin{tabular}{|c|c|c|c|c|}
\hline
$a$& \multicolumn{4}{c|}{${\sigma}/{10^{-41}\mbox{cm}^{2}} $}\\
\hline
&$\nu_e$&$\tilde{\nu}_e$&$\sum_x\nu_x$&Total\\
&T=3.5 MeV&T=5 MeV&T=8 MeV&\\
\hline
 0 & 6.861 & 13.456 & 28.232&48.548 \\
 0.75 & 7.277 & 14.258 & 29.747&51.281 \\
 1.50 & 7.934 & 15.515 & 32.049&55.497 \\
 2.00 & 8.531 & 16.649 & 34.049 &59.229\\
 3.00 & 10.150 & 19.683 & 39.021&68.854 \\
 4.00 & 12.364 & 23.708 & 44.772&80.844 \\
 5.00 & 15.176 & 28.568 & 50.537&94.281\\
 \hline
\end{tabular}
\end{center}
\end{table}
\begin{figure}
 \begin{center}
 \subfloat[]
{
\rotatebox{90}{\hspace{-0.0cm}{$\frac{\sigma(E_{th})}{\sigma(0)}\rightarrow$}}
\includegraphics[scale=0.6]{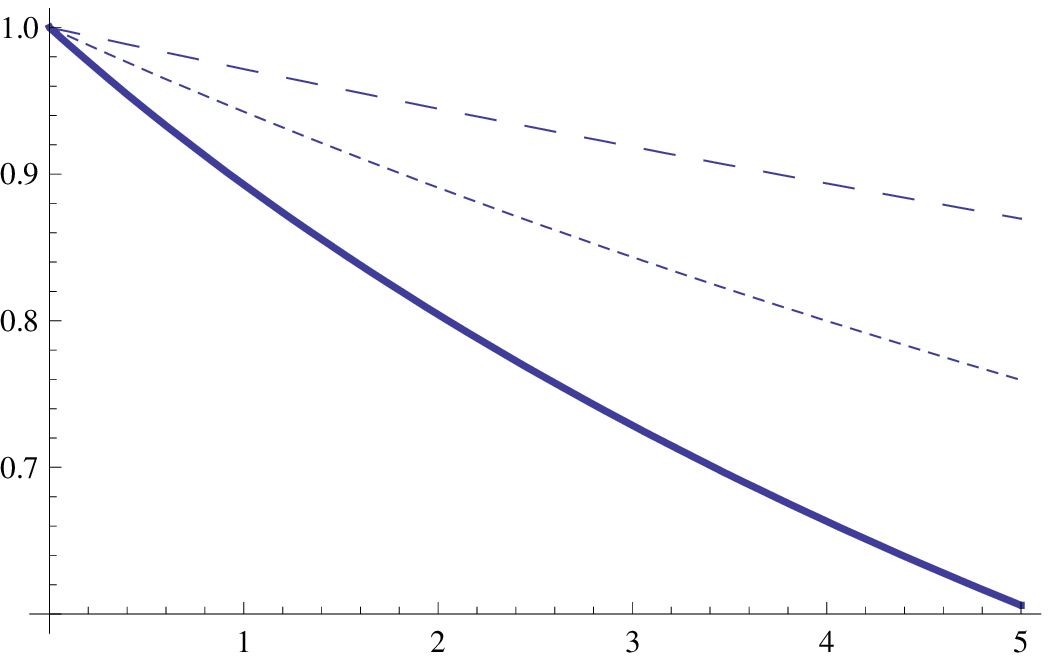}
}\\
 \subfloat[]
 {
\includegraphics[scale=0.6]{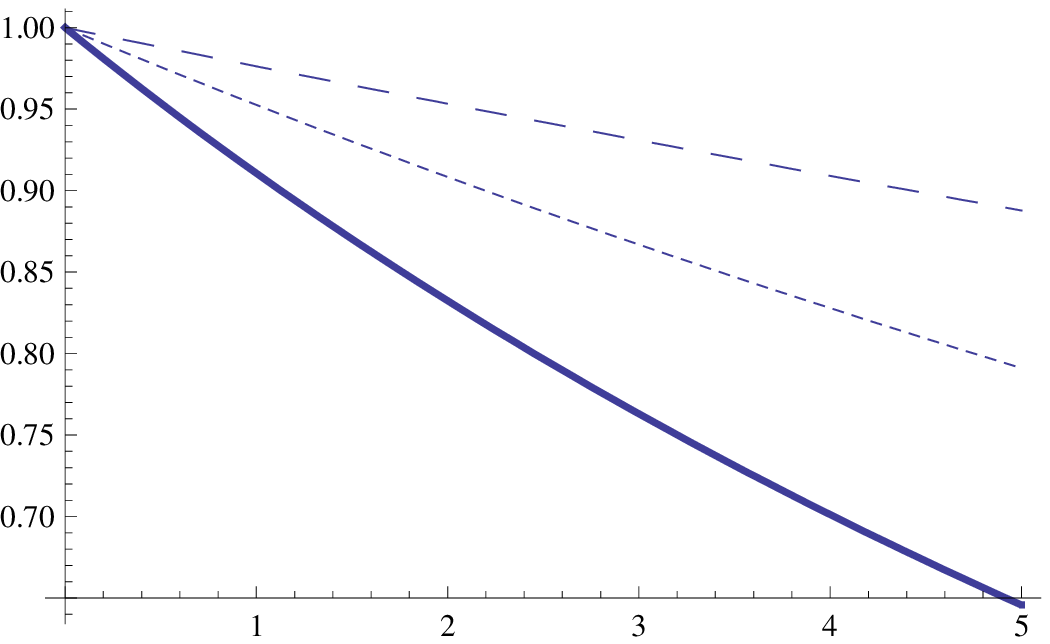}
}
\\
{\hspace{-0.0cm} {$E_{th}\rightarrow $keV}}
 \caption{The same as in Fig. \ref{Fig:tsigma.131} in the case of the Ne target.}
 \label{Fig:tsigma.20}
  \end{center}
  \end{figure}
  
   \begin{figure}[!ht]
 \begin{center}
 \subfloat[]
{
\rotatebox{90}{\hspace{-0.0cm}{$\frac{\sigma(E_{th})}{\sigma(0)}\rightarrow$}}
\includegraphics[scale=0.7]{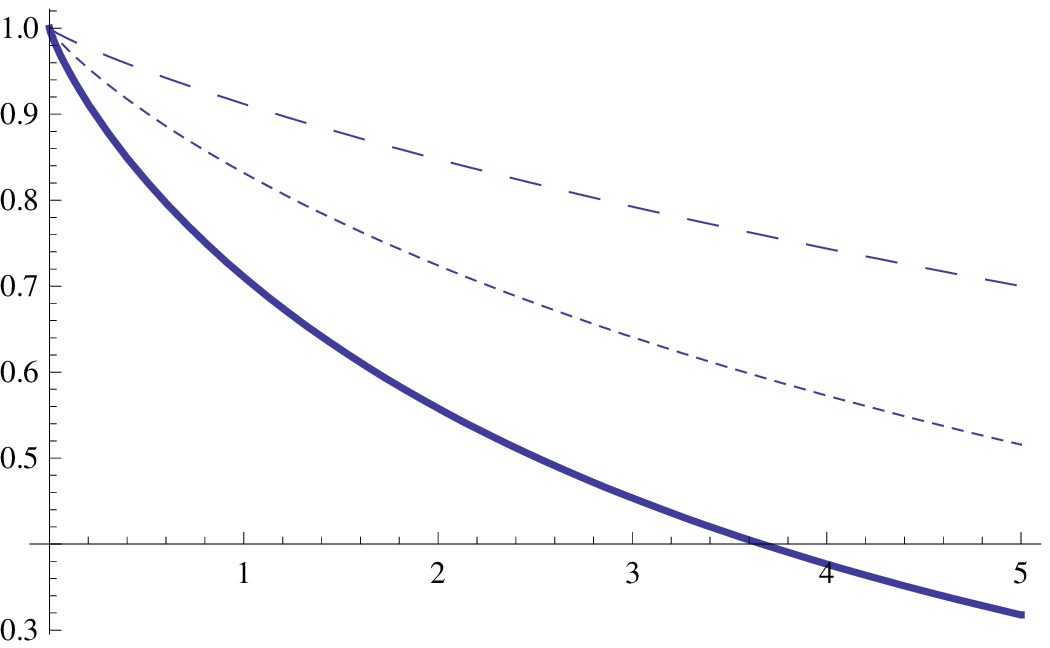}
}\\
 \subfloat[]
 {
\includegraphics[scale=0.6]{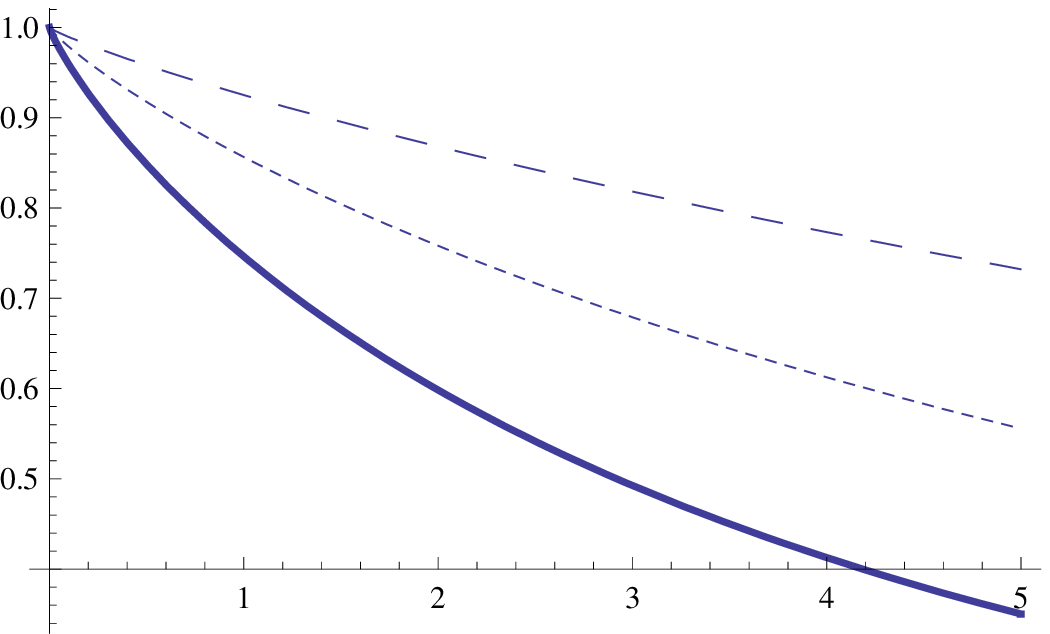}
}
\\
{\hspace{-0.0cm} {$E_{th}\rightarrow $keV}}
 \caption{The same as in Fig. \ref{Fig:qtsigma.131} in the case of the Ne target.}
 \label{Fig:qtsigma.20}
  \end{center}
  \end{figure}
  
  \begin{table}[t]
\caption{ The same as in Table \ref{tab:rate.131} in the case of the target Ne.}
\label{tab:rate.20}
\begin{center}
\begin{tabular}{|r|r|r|}
\hline
a&R=10m&R=4m\\
\hline
  0 & 160 & 10 \\
 0.75 & 163 & 10 \\
 1.50 & 167 & 11 \\
 2.00 & 170 & 11 \\
 3.00 & 177 & 11 \\
 4.00& 185 & 12 \\
 5.00 & 190 & 12\\
  \hline
\end{tabular}
\end{center}
\end{table}
In the presence of a detector threshold of even  1 keV the above rates are reduced by about 5$\%$ (10$\%$ ) in the absence (presence) of quenching.
\section{Terrestrial sources with similar spectrum}
The spherical TPC detector can become a dedicated SN neutrino spectrum. Untill such SN explosion takes place it can be tested using terrestrial neutrino sources with spectra similar to those of SN neutrinos. Two such possibilities come to mind:
\subsection{ The Oak Ridge  Spallation Neutron Source (SNS)}
\begin{figure}[!ht]
 \begin{center}
\includegraphics[scale=0.8]{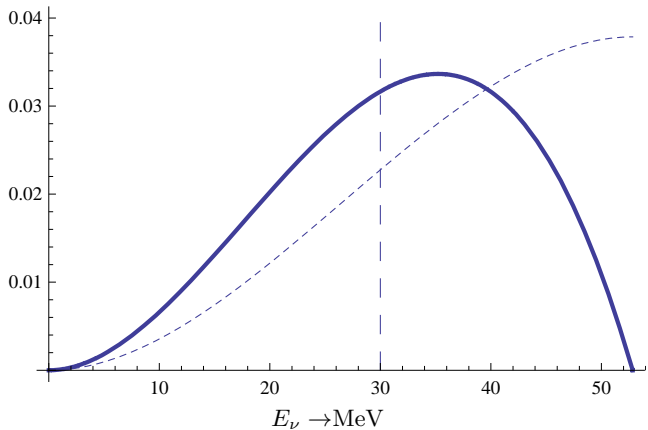}
 {\hspace{1.0cm} {$E_{\nu} \rightarrow$MeV}}
 \caption{The  neutrino spectrum from stopped pions. The normalized solid  and dotted curves correspond
to $\nu_e$ and $\tilde{\nu}_{\mu}$ respectively.  Also shown is the discreet $\nu_{\mu}$ spectrum (dashed vertical line).}
 \label{fig:nuspec}
 \end{center}
  \end{figure}

\begin{figure}[!ht]
 \begin{center}
 \subfloat[]
 {
 \rotatebox{90}{\hspace{-0.0cm}{$\frac{d \sigma}{d T_A}\rightarrow 10^{-38}$}cm$^2/$keV}
\includegraphics[scale=0.7]{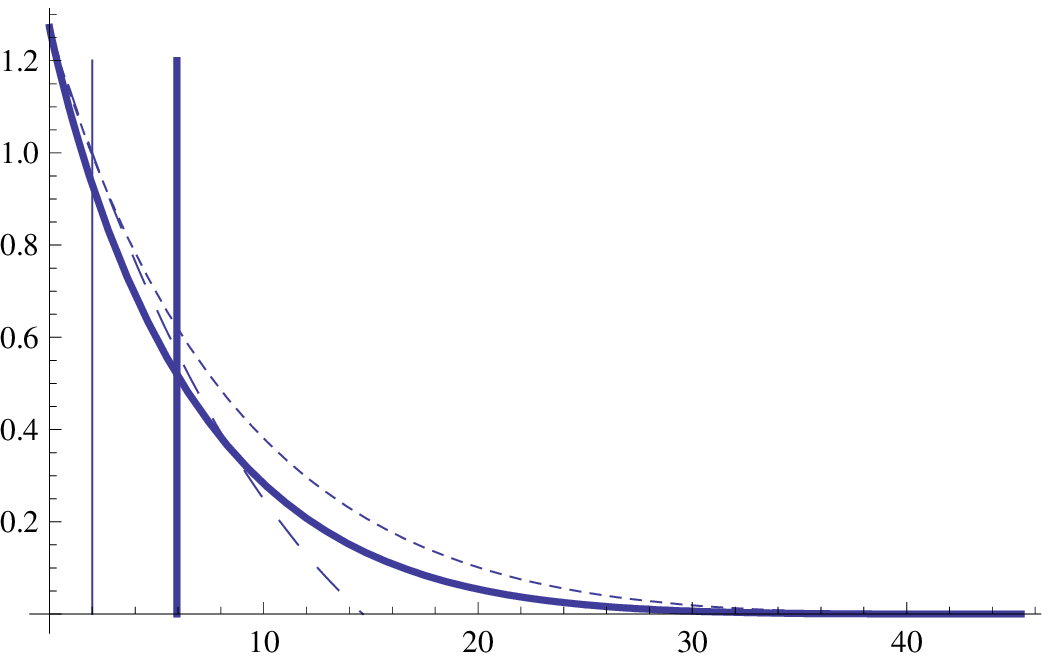}
}\\
\subfloat[]
{
 \rotatebox{90}{\hspace{-0.0cm}{$\frac{d \sigma}{d T_A}\rightarrow 10^{-40}$}cm$^2/$keV}
\includegraphics[scale=0.7]{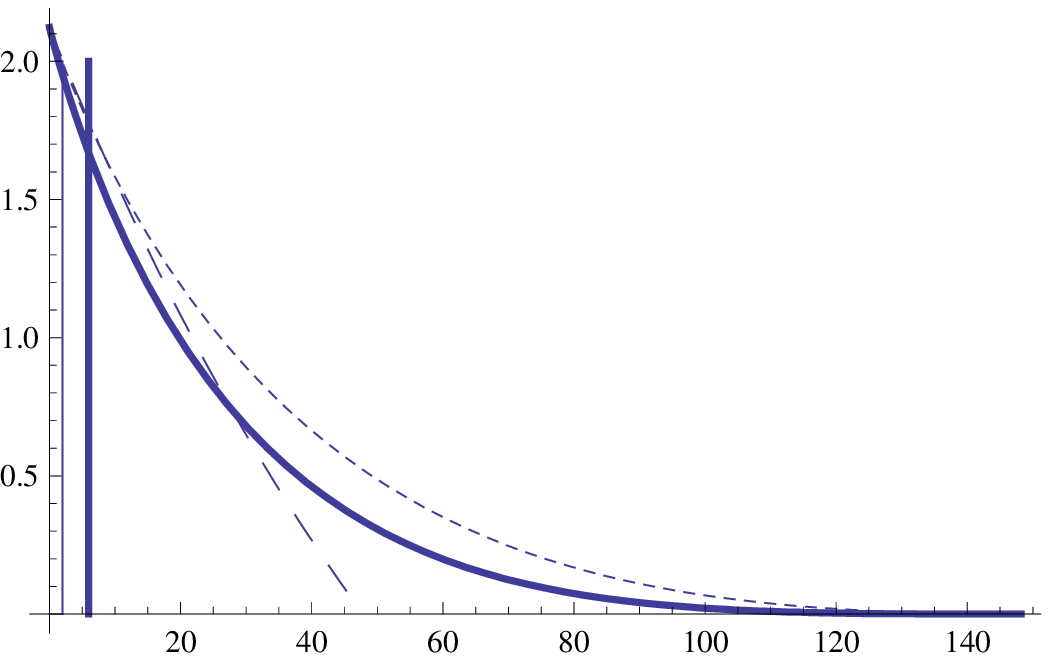}
}\\
{\hspace{1.0cm} {$T_a \rightarrow$keV}}
   \caption{The differential neutrino nucleus cross section, in the case of SNS source, as a function of the recoil energy in kev for the A=131 target (a) and A=40 target (b). The  notation is the same as that of the curves  of Fig. \ref{fig:nuspec}.
The fine vertical line corresponds to a threshold of 2 keV. The phase space on its left
is unavailable. Due to the presence of quenching the  excluded phase space becomes larger (on the left of the thick line), which leads to a smaller total rate.}
 \label{fig:dsigmadT131}
 \end{center}
  \end{figure}

\begin{figure}[!ht]
 \begin{center}
 \subfloat[]
{
\rotatebox{90}{\hspace{-0.0cm}{$\frac{\sigma(E_{th})}{\sigma(0)}\rightarrow$}}
\includegraphics[scale=0.7]{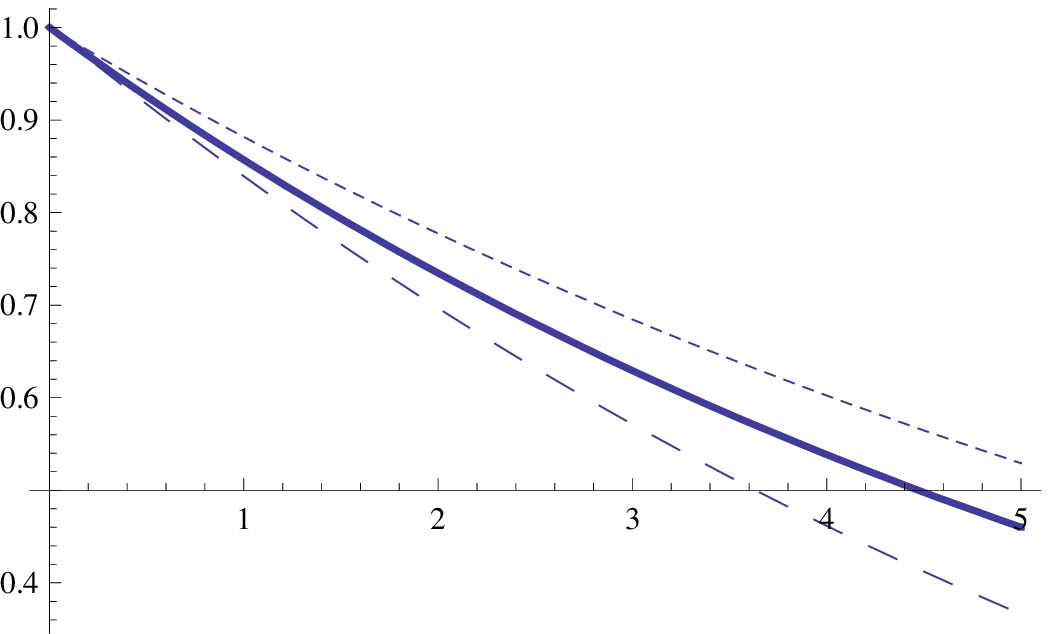}
}\\
 \subfloat[]
 {
\includegraphics[scale=0.7]{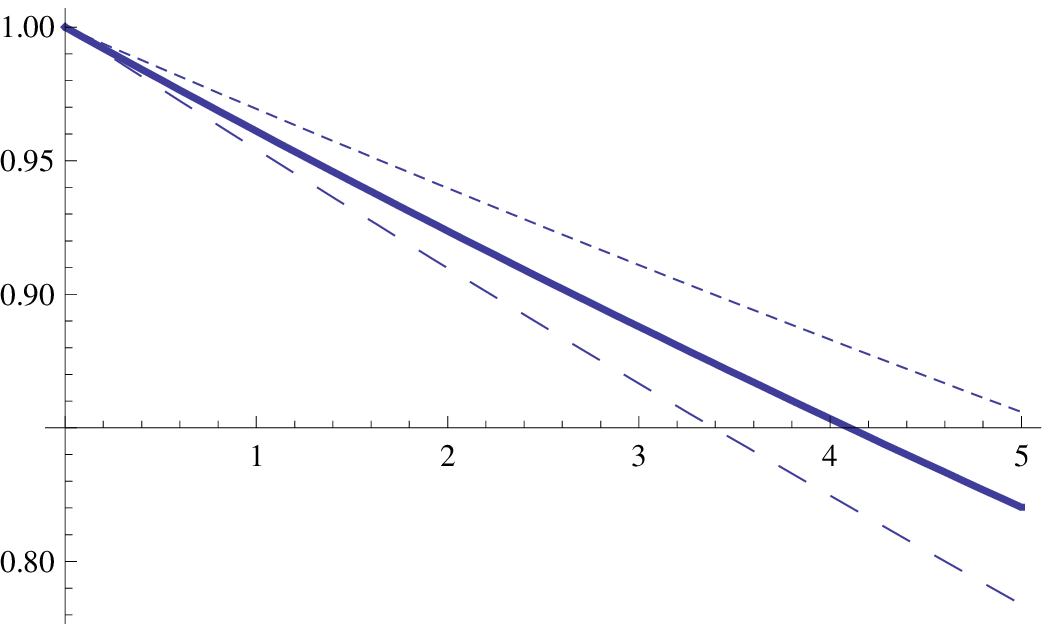}
}
\\
{\hspace{-0.0cm} {$E_{th}\rightarrow $keV}}\\
 \caption{The ratio of the cross section at threshold $E_{th}$ divided by that at zero threshold as a function of the threshold energy in keV, corresponding to A=131 (a) and A=40 (b).
  Otherwise the notation is the same as in Fig. \ref{fig:dsigmadT131}}
 \label{Fig:tsigma.SNS}
  \end{center}
  \end{figure}

\begin{figure}[!ht]
 \begin{center}
 \subfloat[]
{
\rotatebox{90}{\hspace{-0.0cm}{$\frac{\sigma(E_{th})}{\sigma(0)}\rightarrow$}}
\includegraphics[scale=0.7]{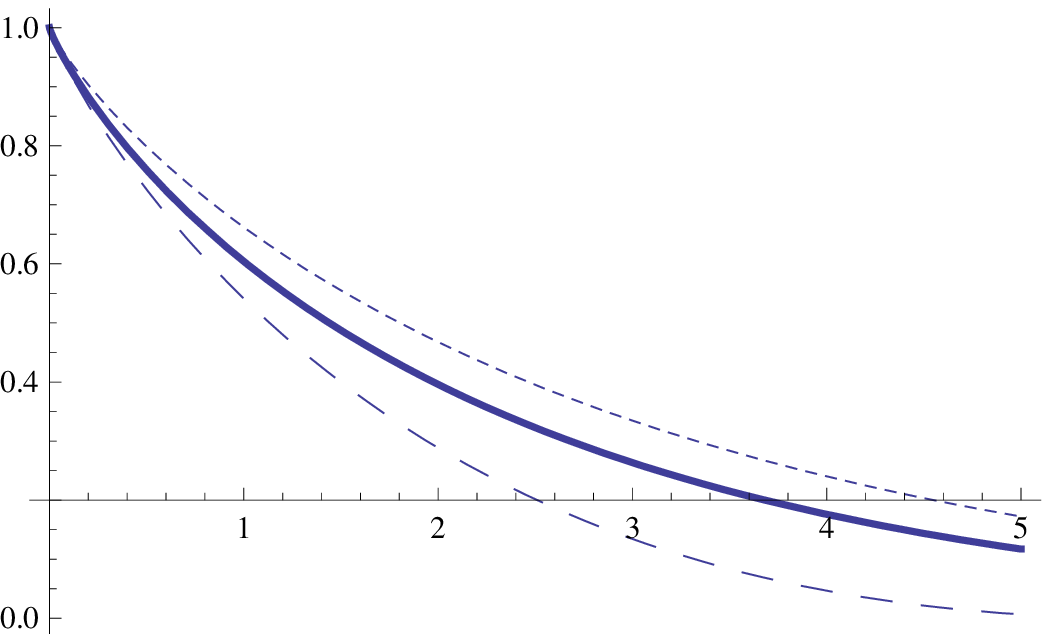}
}\\
 \subfloat[]
 {
\includegraphics[scale=0.7]{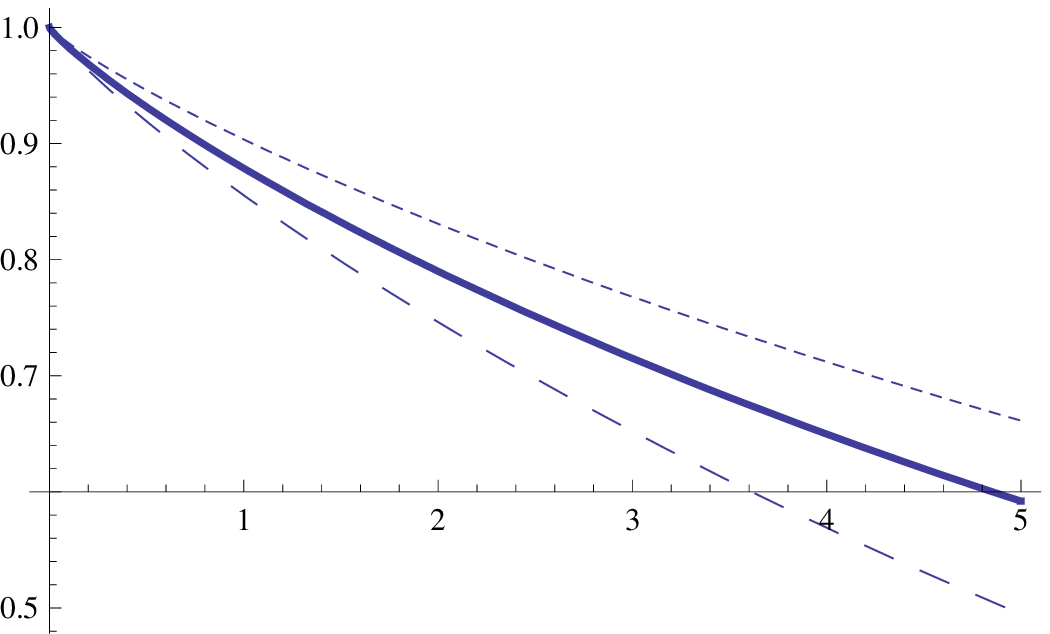}
}
\\
{\hspace{-0.0cm} {$E_{th}\rightarrow $keV}}\\
 \caption{The same as in Fig. \ref{Fig:tsigma.SNS} in the presence of quenching.}
 \label{Fig:qtsigma.SNS}
  \end{center}
  \end{figure}

 SNS is a prolific pulsed source of  
electron and muon neutrinos as well as muon antineutrinos \cite{AVIGNONE},\cite{VerAvGiom09}.
The normalized neutrino spectra can be described very well by the shapes exhibited in
Fig. \ref{fig:nuspec}
  The differential cross sections obtained after folding with the neutrino spectra are shown in  Fig. \ref{fig:dsigmadT131}.
   Integrating these differential cross sections
  we obtain for A=131 at zero threshold
 the total cross sections:
 $$
 (6.21,19.1,7.47)\times 10^{-38}\mbox{cm}^2
  $$ 
 for $\nu_e$, $\tilde{\nu}_{\mu}$ and $\nu_{\mu}$ respectively. In the case of the A=40 we find
  $$
  (5.33,6.83,4.58)\times 10^{-39}\mbox{cm}^2.
  $$
  The effect of threshold is exhibited in Figs \ref{Fig:tsigma.SNS}-\ref{Fig:qtsigma.SNS}.
  
\begin{figure}[!ht]
 \begin{center}
 \rotatebox{90}{\hspace{1.0cm} {$s(R/L) \rightarrow$}}
\includegraphics[scale=0.7]{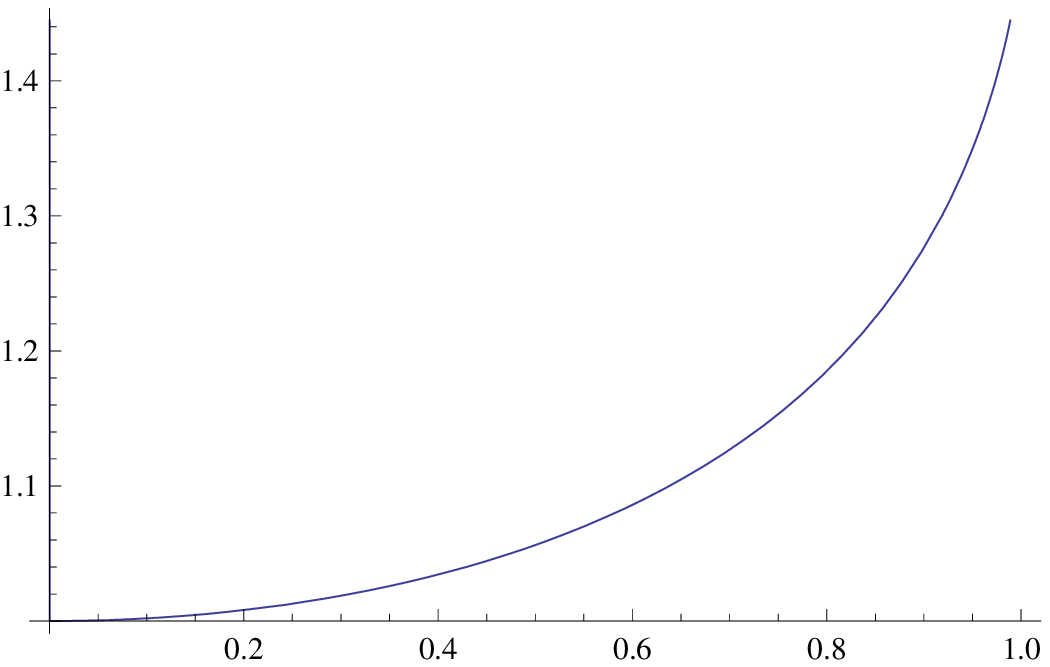}\\
{\hspace{1.0cm} {$\frac{R}{L} \rightarrow$}}
 \caption{The parameter $s(R/L)$ (see text) in the case of a sphere of radius $R$ whose center is at a distance $L>R$ from the source.}
 \label{Fig:sphere}
 \end{center}
  \end{figure}

  The SNS facility provides an excellent opportunity to employ and test the spherical TPC gaseous detector \cite{sphere08}. There remain, however, some experimental issues that need be resolved first. These have do with the fact that the neutrino flux is changing from point to point inside the detector.  The number of expected events for a vessel of volume $V$ under pressure $P$ and temperature $T$  takes the form:
\beq
 {\cal R}= 3.156\times 10^7 \frac{t}{1y}\Phi(\nu,L)\sigma(A,N) \frac{PV}{kT} s(V,L) 
 \eeq
 where the parameter $s(V,L)$ is a geometrical factor needed when a large detector is close to the source. It  depends on the shape of the vessel and the distance $L$ of its geometric center  from the source.
  In the case of  sphere of radius R with its center at a distance $L$ from the source he function s(V,R) depends only on the ratio $R/L$, i.e. $s(V,L)\rightarrow s(R/L)$, and it is given by
\barr
s(R/L)&=&\frac{L^2}{(4 /3)\pi R^3} 2 \pi L \int_0^{R/L} x^2 dx \nonumber\\\int_ 0^{\pi } d \theta &&\sin{\theta}\frac{1}{1+x^2+2 x \cos{\theta}},\quad x=\frac{r}{L}
\earr
The above expression takes into account the fact that, if the sphere is close to the source, the neutrino flux is changing from point to point inside the sphere. Spherical coordinates $(r,\theta,\phi)$, are used to specify any point inside the sphere. The origin of coordinates was chosen at the center of the sphere with  polar axis  the straight line from the source to the center of the sphere. With the above  choice the flux is independent of the angle $\phi$. The function $s(R/L)$ is given by:
\beq
s(x)=\frac{3}{2x^3}\left ( \frac{x^2-1}{2}\ln{\frac{1+x}{1-x}}+x \right )
\eeq
Its behavior is exhibited in Fig. \ref{Fig:sphere}. The geometric factor for the actual experimental set up, $L>>R$, is close to unity.
Using 
$$ \Phi(\nu,L=50\mbox{m})=1.95\times 10^{6}\mbox{cm}^{-2}\mbox{s}^{-1}$$
for each flavor we obtain
  the data of table \ref{table:totalrates2}.
 \begin{table}[t]
\caption{The number of events in a year originating from the Spallation Neutron Source (SNS) for nuclear recoils due
to neutral current neutrino-nucleus scattering as a function of $(P,R,L)$ where $P$ is the pressure of the gas in the vessel, $R$ is the vessel radius and $L$ is the distance from the source. The detector is  at temperature $T=300~^0$K and is assumed to operate with zero threshold.
}
\label{table:totalrates2}
\begin{center}
\begin{tabular}{|c|c|c|c|c|}
\hline
target&distance&\multicolumn{3}{c|}{$P=$10Atm}\\
\hline
& &  $R=$10m&$R$=5m&$R$=1.3m\\
\hline
& $L$=50m&  $1.85\times 10^8 $& $2.30\times 10^7$ & $4.03\times 10^5$\\
 $^{131}_{77}$Xe&$L$=100m&$4.60\times 10^7 $& $5.74\times 10^6$ & $1.00\times 10^5$  \\
 &$L$=150m &$2.04\times 10^7$ & $2.55\times 10^6$ &  $4.48 \times 10^4$\\
 \hline
  &$L=$50m& $9.40\times 10^6$ & $1.17\times 10^6 $& $2.05 \times 10^4$ \\
 $^{40}_{18}$Ar&$L$=100m&$ 2.34\times 10^6$ & $2.92\times 10^5$ & $5.12\times 10^3$ \\
 &$L$=150m   &$1.04\times 10^6$ & $1.30\times 10^5$ & $2.28\times 10^3$\\
   \hline
\end{tabular}
\end{center}
\end{table}
  \subsection{Boosted Radioactive Sources}
\begin{figure}[!ht]
 \begin{center}
 \subfloat[]
 {
\includegraphics[scale=0.6]{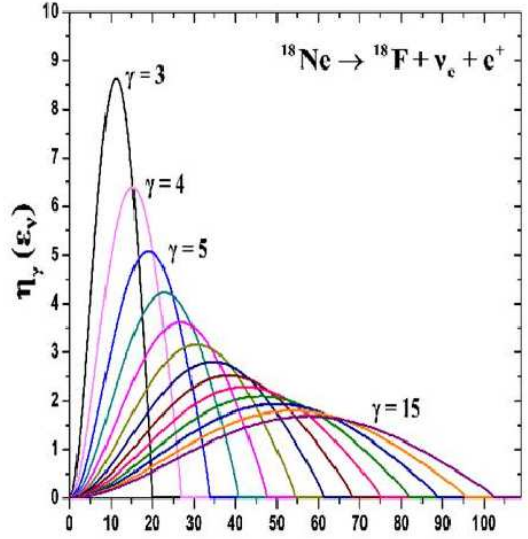}
}\\
\subfloat[]
{
\includegraphics[scale=0.6]{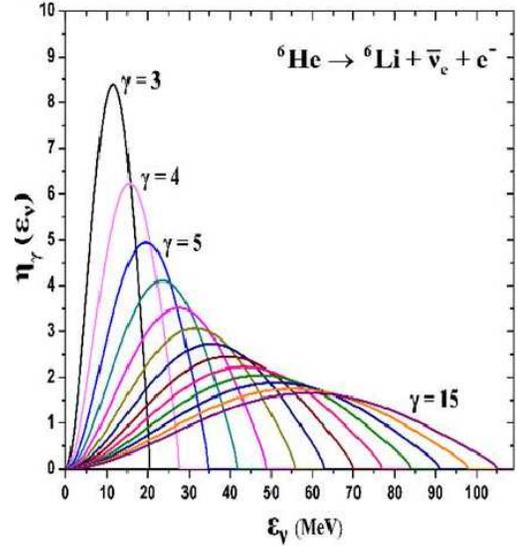}
}
 \caption{The spectrum of a typical boosted neutrino source (a) and antineutrino (b) for various values of the boosting parameter $\gamma$.}

 \label{Fig:radsource}
 \end{center}
  \end{figure}

  In recent years in the process of developing neutrino factories it has become feasible to obtain radioactive sources \cite{Zucchelli02} for both electron neutrinos and antineutrinos. There exist many candidate  sources, which can be boosted up to high energies, with boosting factors $\gamma=E_b/M_bc^2$ ranging up to 20,
  yielding neutrino spectra not very different from those expected for SN neutrinos\cite{AUTIN03}. These sources can live long enough ($\tau=\gamma \tau_0$) so that experiments testing the SN detectors can be performed.
  The spectra of two popular such sources, whose advantages have already been  discussed \cite{AUTIN03}, are shown in Fig. \ref{Fig:radsource}.
\section{Conclusions}
From the above results one can clearly see the advantages of a gaseous spherical TPC detector dedicated for SN neutrino detection. 
The first idea is to employ a small size spherical TPC detector filled with a high
pressure noble gas.  An enhancement of the neutral current component is achieved via the coherent
effect of all neutrons in the target. Thus employing, e.g., Xe at $10$ Atm, with a feasible threshold energy
of about $100$ eV in  the recoiling nuclei,
 one expects between $700$ and $900$ events for a sphere of radius 5m. Employing $^{40}$Ar  one expects between 200 and 300 events but with a vessel of larger radius (R=10m). If necessary, this detector can be tested with Earth neutrino sources, which have a neutrino spectrum analogous to that of a SN.
 
 This  detector, dedicated for SN detection, should not be viewed as a competitor of the huge constructions currently  being envisioned for various purposes including supernova neutrino detection \cite{LLDAPP07}\footnote{ See, e.g.  in the Proceedings of the FIFTH SYMPOSIUM ON LARGE TPCs
FOR LOW ENERGY RARE EVENT DETECTION
and workshop on neutrinos from Supernovae.:  T. Patzak, Laguna: future Megaton detectors in Europe; A Curioni, The GLACIER project and related R and D;
K. Scholberg, "Supernova Neutrino Detection in Water Cherenkov Detectors"
; A. Ianni, "Supernova Neutrino Detection With Liquid Scintillators"
; L. Koepke, "Supernova Neutrino Detection with IceCube"'; I. Gil-Bottella,  "Supernova neutrino detection in LAr TPCs"
.}. The information provided by the neutral current detectors, which are not sensitive to neutrino oscillation effects, will provide information about the primary supernova neutrino flux and aid the analysis of detectors exploiting the charged current interaction. It can also serve as an insurance, if anything goes wrong with the large detectors.
 
The second idea  is to  build several such low cost and robust detectors and install them in 
several places over the world.  First estimates show that the required background level
is modest and therefore there is no need for a deep underground laboratory. A mere  100 meter 
water equivalent
coverage seems to be sufficient to reduce the cosmic muon flux at the required level 
(in the case of many such detectors in coincidence, a modest shield is sufficient). 
The maintenance of such systems, quite simple and needed only once every  few years, could be easily assured by universities or even by 
secondary schools, with only 
 specific running programs. 
Admittedly  such a detector scheme, measuring low energy nuclear recoils from neutrino 
nucleus elastic scattering, will not be able to  determine the incident neutrino vector and, therefore, 
it is not possible  to localize  the supernova this way. This can be achieved by a cluster of such detectors in 
coincidence by a triangulation technique. \\
 A network of such detectors in coincidence with a sub-keV threshold could also be used to observe
 unexpected low energy events. This low energy range has never been explored using massive
 detectors. A challenge of great importance will be the synchronization of such a detector cluster
with the astronomical $\gamma$-ray burst telescopes to establish whether low energy recoils are
emitted in coincidence with the mysterious $\gamma$ bursts. \\ 
 In summary: networks of  such dedicated gaseous TPC detectors, made out of simple, robust and cheap technology,
 can be simply managed by an international scientific consortium and operated by students. This network
 comprises a system, which can be cheaply maintained
for several decades (or even centuries). Obviously this is   is a key point towards preparing to observe
 few galactic supernova explosions.

\section*{Acknowledgement}
The author would like to express his appreciation to the organizers of the   
"FIFTH SYMPOSIUM ON LARGE TPCs
FOR LOW ENERGY RARE EVENT DETECTION
and workshop on neutrinos from Supernovae" for their financial support and hospitality.
\section*{References}

\end{document}